\documentclass[journal,twocolumn]{IEEEtran}

\usepackage{amsmath}
\usepackage{cite}
\usepackage{epsf}

\newtheorem{assumption}{Assumption}
\newtheorem{identity}{Identity}

\def\vec#1{{\bf #1}}

\def\Tr#1{\mbox{Tr}\left\{ #1 \right\} }
\def\outer{\otimes}

\def\bG{{\bf G}}
\def\bGhat{{\hat{\bf G}}}
\def\nt{{n_t}}
\def\nr{{n_r}}
\def\bRcal{{\bf{\cal  R}}}
\def\bTcal{{\bf{\cal  T}}}
\def\dRcal{\mbox{\boldmath$\delta$}{\bf {\cal R}}}
\def\dTcal{\mbox{\boldmath$\delta$}{\bf {\cal T}}}

\begin{document}
\bstctlcite{BSTcontrol}

\title{On the Outage Capacity of Correlated Multiple-Path MIMO Channels}

\author{ Aris L. Moustakas and Steven H. Simon%
\thanks{A. L. Moustakas
(email: arislm@phys.uoa.gr), and S. H. Simon (email:
shsimon@bell-labs.com)}}

\maketitle
\date{}

\begin{abstract}
The use of multi-antenna arrays in both transmission and reception
has been shown to dramatically increase the throughput of wireless
communication systems. As a result there has been considerable
interest in characterizing the ergodic average of the mutual
information for realistic correlated channels. Here, an approach
is presented that provides analytic expressions not only for the
average, but also the higher cumulant moments of the distribution
of the mutual information for zero-mean Gaussian MIMO channels
with the most general multipath covariance matrices when the
channel is known at the receiver. These channels include multi-tap
delay paths, as well as general channels with covariance matrices
that cannot be written as a Kronecker product, such as
dual-polarized antenna arrays with general correlations at both
transmitter and receiver ends. The mathematical methods are
formally valid for large antenna numbers, in which limit it is
shown that all higher cumulant moments of the distribution, other
than the first two scale to zero. Thus, it is confirmed that the
distribution of the mutual information tends to a Gaussian, which
enables one to calculate the outage capacity. These results are
quite accurate even in the case of a few antennas, which makes
this approach applicable to realistic situations.
\end{abstract}


\section{Introduction}

\PARstart{F}{ollowing} pioneering work by
\cite{Foschini1998_BLAST1, Telatar1995_BLAST1} it has become clear
that the use of multi-antenna arrays in transmission and reception
can lead to significantly increased bit-rates. This has led to a
flurry of work calculating the narrowband ergodic mutual
information of such systems, i.e. the mutual information averaged
over  realizations of the channel, using a variety of channel
models and analytic techniques. For example, the ergodic capacity
was calculated asymptotically for a large number of
antennas,\cite{Rapajic2000_InfoCapacityOfARandomSignatureMIMOChannel,
Moustakas2000_BLAST1, Chuah2002_MIMO1, Lozano2002_MIMO1,
Muller2002_RandomMatrixMIMO, Sengupta2000_BLAST1,
Mestre2003_CapacityOfMIMOChannelsAsymptoticEvaluationUnderCorrelatedFading,
Tulino2004_INDLargeNCapacity, Skipetrov2003_MIMO,
Tulino2004_RMTInfoTheoryReview} or for large and small
\cite{Lozano2003_LowPowerMIMOCapacity,
Tulino2004_INDLargeNCapacity} signal-to-noise ratios, using a
variety of assumptions for the statistics
\cite{Moustakas2003_MIMO1, Tulino2004_INDLargeNCapacity} of the
fading channel.

To better understand the characteristics of realistic information
transmission through fading channels, it is important to analyze
the full distribution of the mutual information over realizations
of fading. For example, the outage capacity
\cite{Ozarow1994_OutageCapacity} is sometimes a more realistic
measure of capacity for delay constrained fading channels. In
addition, the distribution of the mutual information provides
information about the available diversity in the system
\cite{Zheng2003_DiversityMultiplexing}: the smaller the variance,
the lower the probability of outage error when transmitting at a
fixed rate. Finally, having an analytic expression for the
distribution of the mutual information allows one to simulate a
system of multiple users in a simple
way.\cite{Hochwald2002_MultiAntennaChannelHardening} Recently,
\cite{Sengupta2000_BLAST1, Moustakas2003_MIMO1} analytically
calculated the first few moments of the distribution of the
narrowband mutual information, asymptotically for large antenna
numbers with spatial correlations. This analysis showed that the
distribution is approximately Gaussian even for a few antennas,
also seen in
\cite{Smith2002_OnTheGaussianApproximationToTheCapacityOfWirelessMIMOSystems,
Hochwald2002_MultiAntennaChannelHardening}. More recently, other
methods were devised to calculate all moments of the mutual
information distribution exactly for some channel types.
\cite{Smith2003_CapacityMIMOsystems_with_semicorrelated_flat_fading,
Kang2002_CapacityOfMIMORicianChannels,
Chiani2003_CapacityOfSpatiallyCorrelatedMIMOChannels,
Kiessling2004_ExactMutualInfoCorrelatedMIMOChannels,
Simon2004_EigenvalueDensityOfCorrRandomWishartMatrices,
Simon2004_CapacityOfCorrRandomWishartMatrices} Also,
\cite{Tulino2004_INDLargeNCapacity} calculated the ergodic mutual
information in the large antenna limit for independent
non-identically distributed (IND) channels, and extended their
results to correlated channels with special restrictions on the
correlations of different paths.

The above literature did not analyze the statistics of the mutual
information for Gaussian channels with general
non-Kronecker-product correlations.
\cite{Oestges2003_ImpactDiagonalCorrelationsMIMO,
Orcelik2003_KroneckerProductDeficiencies, SCM_3GPP_TR25_996,
Shafi2005_3DPolarizedMIMOChannels} These types of channels are
becoming increasingly important to study, as it has recently been
proposed that they appear in several situations, such as channels
for generally correlated antennas with multiple polarizations.
\cite{SCM_3GPP_TR25_996, Shafi2005_3DPolarizedMIMOChannels}

Furthermore, the above works have generally focused on the case of
narrowband flat-fading channels. However, the use of wide-band
signals with non-trivial resolvable multipath necessitates the
analysis of the mutual information in the presence of multipath.
\cite{Bolcskei2002_OFDMMIMOCapacity,
Oyman2003_StatPropertiesMutInfoMIMO} showed that the capacity of
the wideband channel depends only on narrowband quantities, such
as total average power etc. Subsequently, other authors have
analyzed the wideband ergodic capacity using asymptotic methods.
\cite{Muller2002_RandomMatrixMIMO,
Liu2003_CapacityScalingWidebandCorrMIMOChannels} In a first
attempt to describe the wideband distribution of the mutual
information,
\cite{Barriac2005_OutageRatesSpaceTimeCommunicationsWidebandChannels}
 suggested that
the distribution is Gaussian, if the number of independent paths
is large. However, in many instances of interest the number of
paths seen is small. \cite{Pedersen2000_SpaceTimeChannelModel,
SCM_3GPP_TR25_996} It would thus be useful to analyze the effects
of multi-path on the wideband mutual information of Gaussian MIMO
fading channels of {\em arbitrary} multipath behavior in an
analytic fashion. Although the exact methods mentioned above
\cite{Smith2003_CapacityMIMOsystems_with_semicorrelated_flat_fading,
Kang2002_CapacityOfMIMORicianChannels,
Chiani2003_CapacityOfSpatiallyCorrelatedMIMOChannels,
Kiessling2004_ExactMutualInfoCorrelatedMIMOChannels,
Simon2004_EigenvalueDensityOfCorrRandomWishartMatrices,
Simon2004_CapacityOfCorrRandomWishartMatrices} can calculate all
moments of the distribution for narrowband channels, they cannot
be generalized to multi-path channels. Therefore, to make
progress, one needs to rely on asymptotic methods.

In this paper we extend work done in \cite{Moustakas2003_MIMO1} to
provide analytic expressions for the statistics of the mutual
information in the presence of multi-path with general spatially
correlated channels. We assume that the instantaneous fading
channel is known to the receiver but not the transmitter. Our
results generalize the mutual information results of
\cite{Tulino2004_INDLargeNCapacity} for Gaussian channels to
arbitrary zero-mean Gaussian correlated channels. The
 paths may or may not have the same delay. The methods
used here apply the concept of replicas, which was initially
introduced in statistical physics for understanding random systems
\cite{Edwards1975_FirstReplicaPaper}, but  in recent years have
seen several applications in information theory.
\cite{Moustakas2000_BLAST1, Montanari2000_TurboCodesPhaseTrans,
Tanaka2002_ReplicasInCDMAMUD,
Guo2003_ReplicaAnalysisOfLargeCDMASystems,
Muller2002_RandomMatrixMIMO}.

In particular, we obtain the following results:
\begin{itemize}
\item We use the replica method to calculate the moment generating function
 of the mutual information, averaging over general multipath, non-Kronecker product channels.
Using this approach we derive expressions for its first three
moments (mean, variance and skewness). As in
\cite{Moustakas2003_MIMO1} we find that for large antenna numbers
$n$, the average of the distribution is of order $n$,  the second
moment of the distribution is of order unity, and the third moment
is order $1/n$ respectively, while all other moments scale with
higher powers of $1/n$. Thus, for large $n$ the mutual information
distribution approaches a Gaussian. Therefore, the outage mutual
information can be expressed simply in terms of the mean and the
variance of the distribution (section \ref{sec:math_framework}).

\item We optimize the mean mutual information with respect to the input signal distribution to obtain the ergodic
capacity (section \ref{sec:ergodic_capacity}).

\item We demonstrate the dependence of the whole distribution of the mutual information on
the specifics of the channel by calculating the mean and variance
of the mutual information for a number of simple multipath
channels.

\item We also compare these Gaussian distributions with numerically generated ones and find
very good agreement, even for a few antennas. This validates the
analytical approach presented here for use in realistic situations
with small antenna numbers.

\end{itemize}

\subsection{Outline}
\label{sec:outline}

In the remainder of this section we define relevant notation. In
section \ref{sec:Multipath fading MIMO channel model} we describe
the MIMO channels for which our method is applicable, in both the
temporal and the frequency domain. In section \ref{sec:Mutual
Information} we define the wideband mutual information and in
section \ref{sec:Statistics of Mutual Information} the statistics
of its distribution. Subsequently, in section
\ref{sec:math_framework} the mathematical framework of the method
to calculate the generating function of the mutual information is
presented. Also, the calculation of the ergodic capacity (section
\ref{sec:ergodic_capacity}), its variance (section
\ref{sec:variance_mutual_info_iid}) and the higher order moments
of the distribution (section \ref{sec:higher_order_terms}) are
discussed. Section \ref{sec:special case: R_l independent of l}
deals with a alternative derivation of the results for the case
when the receive correlation matrix is the same for all paths,
while section \ref{sec:Special Case 2: Narrowband Multipath}
briefly discusses the case of narrowband multipath, where all
paths arrive at the same delay tap. In section \ref{sec:Analysis
of Results} a few specific cases are analyzed analytically and
compared to numerical Monte-Carlo calculations. Appendix
\ref{app:complex_integrals} summarizes a number of complex
integral identities employed in the main section, while Appendices
\ref{app:Derivation of gnu_S} and \ref{app:Details for Saddle
Point Analysis of gnu_S} contain some details for various steps in
section \ref{sec:math_framework}. Appendix
\ref{app:higher_order_terms} includes some guiding details of the
calculation of the higher order terms in section
\ref{sec:higher_order_terms}. Finally, Appendix \ref{app:optimalQ}
describes the procedure of evaluating the capacity-achieving
transmission covariance $\vec Q$.

\subsection{Notation}
\label{sec:notation} %

\subsubsection{Vectors/Matrices}
Throughout this paper, we will use bold-faced upper-case letters
to denote matrices, e.g. $\vec X$, with elements given by
$X_{ab}$, bold-faced lower-case letters for column vectors, e.g.
$\vec x$ with elements $x_a$, and non-bold letters for scalar
quantities. Also the superscripts $T$ and $\dagger$ will indicate
transpose and Hermitian conjugate operations and $\vec I_n$ will
represent the $n$-dimensional identity matrix.

Finally, the superscripts/subscripts $t$ and $r$ will be used for
quantities referring to the transmitter and receiver,
respectively.

\subsubsection{Gaussian Distributions} The real Gaussian distribution with
 zero-mean and unit-variance will be denoted by
${\cal N}(0,1)$, while the corresponding complex, circularly
symmetric Gaussian distribution will be ${\cal CN}(0,1)$.

\subsubsection{Order of Number of Antennas ${\cal O}(n^k)$}
We will be examining quantities in the limit when the number of
transmitters $\nt$ and number of receivers $\nr$, are both large
but their ratios are fixed and finite. We will denote collectively
the order in an expansion over the antenna numbers as ${\cal
O}(n)$, ${\cal O}(1)$, ${\cal O}(1/n)$ etc., irrespective of
whether the particular term involves $\nt$ or $\nr$.

\subsubsection{Integral Measures}
Two general types of integrals over matrix elements will be dealt
with and  the following notation for their corresponding
integration measures will be adopted. In the first type we will be
integrating over the real and imaginary part of the elements of a
complex $m_{rows}\times m_{cols}$ matrix $\vec X$. The integral
measure will be denoted by
\begin{equation}
\label{eq:DcX} %
D\vec X = \prod_{a=1}^{m_{rows}}\prod_{\alpha=1}^{m_{cols}}
\frac{d{\rm Re}\left(X_{a\alpha}\right)d{\rm
Im}\left(X_{a\alpha}\right)}{2\pi}
\end{equation}
The second type of integration is over pairs of complex square
matrices $\bTcal$ and $\bRcal$. Each element of $\bTcal$ and
$\bRcal$ will be integrated over a contour in the complex plane
(to be specified). The corresponding measure will be described as
\begin{equation}
\label{eq:dmuTR1} %
d\mu(\bTcal,\bRcal) =
\prod_{a=1}^{m_{rows}}\prod_{\alpha=1}^{m_{cols}}
\frac{d\bTcal_{a\alpha} d\bRcal_{\alpha a}}{2\pi i}
\end{equation}
In addition, we will define a measure over a set of $L$ pairs of
matrices $\{ \bTcal^l, \bRcal^l \}$ for $l = 0, \ldots, L-1$ to be
given simply by
\begin{equation} \label{eq:dmuTR2}
d\mu( \{ \bTcal^l,\bRcal^l \} ) = \prod_{l=0}^{L-1}
d\mu(\bTcal^l,\bRcal^l)
\end{equation}

\subsubsection{Expectations}We will use the notation $\langle
\, \cdot \, \rangle$ to indicate an expectation over
instantiations of the fading channel.  We will reserve the
notation $E[ \, \cdot\,  ]$ for expectations over transmitted
signals.

\section{Multipath MIMO channel model}
\label{sec:Multipath fading MIMO channel model}

We consider the case of single-user transmission from $\nt$
transmit antennas at a base station to $\nr$ receive antennas at a
mobile terminal over a fading channel with multiple paths with a
finite bandwidth. We assume that the channel coefficients are
known to the receiver, but not to the transmitter. The transmitted
signal can be written in terms of discrete a time series
representing the signals at discrete time steps $m\tau$ for
$m\varepsilon\vec Z$ and $\tau$ the inverse available bandwidth.
Thus we can use the following simple tap-delay
model\cite{Bolcskei2002_OFDMMIMOCapacity, Proakis_book}
\begin{equation}\label{eq:basic_channel_eq}
  \vec y_m = \sum_{l=0}^{L-1}\bG_l \, \vec
  x_{m-m_l} + \vec z_m
\end{equation}
where $\vec x_m$ is the $\nt$-dimensional signal vector
transmitted at time $m\tau$. Similarly, $\vec y_m$ and $\vec z_m$
are the corresponding $\nr$-dimensional received signal and noise
vectors. $\vec z_m$ is assumed an i.i.d. vector with each of its
elements drawn from ${\cal CN}(0,1)$, while $\bG_l$ is the
$\nr\times \nt$-dimensional complex channel matrix at delay times
$m_l\tau$, where $m_l$ is integer-valued. Of course, $\bG_l$ can
be interpreted in a wider sense as an appropriately filtered
version of the channel over the delay interval
$(m_{l-1}\tau,m_l\tau]$. \cite{Proakis_book} Note that in general
all paths need not arrive with different delays, i.e. we have
$m_{l+1}\geq m_l$, with equality when the $l$th and $(l+1)$th
paths arrive within the same delay interval. In fact, all paths
may be assumed to arrive over the same delay interval.

The analysis of multipath channels is simplified considerably by
Fourier-transforming the transmitted and received signal vectors.
In this case the Fourier-transformed received signal is solely a
function of the corresponding Fourier component of the transmitted
signal
\begin{eqnarray}\label{eq:channel_eq_cont_FT}
\hat{\vec y}(\omega) = \hat{\vec G}(\omega) \hat{\vec x}(\omega) +
\hat{\vec z}(\omega)
\end{eqnarray}
where the Fourier transform of the transmitter signal vector
$\hat{\vec x}(\omega)$ is defined by
\begin{eqnarray}\label{eq:FourierTrans_def_infinite}
\hat{\vec x}(\omega) = \sum_{m=-\infty}^{\infty} e^{-i\omega
m\tau} \vec x_m
\end{eqnarray}
with similar definitions for the Fourier components $\hat{\vec
y}(\omega)$, $\hat{\vec z}(\omega)$.  $\hat{\vec G}(\omega)$ is
the Fourier transform of the channel impulse response given by
\begin{eqnarray}\label{eq:FourierTrans_channel}
\hat{\vec G}(\omega) = \sum_{m=0}^{L-1} e^{-i\omega m\tau}\, \bG_m
\end{eqnarray}
Note that (\ref{eq:FourierTrans_def_infinite}) implies that each
symbol vector $\hat{\vec x}(\omega)$ transmitted over a single
frequency is spread over infinite times. As a result, it sees no
interference from other frequency components due to multi-path. In
practice, and in order to avoid mixing between close frequencies
due to Doppler fading, one has to transmit each symbol over a
finite time window, therefore essentially using a discrete set of
frequency components, e.g. $\omega_k = 2\pi k/(M\tau)$, with
$k=0,\ldots,M-1$. The number of discrete frequency components $M$
is usually chosen so that the symbol duration is less than the
coherence time of the channel $t_{coh}$, i.e. $M<t_{coh}/\tau$.
One can then send different symbols one after the other. However,
there is a residual ISI interference due to multipath and the
finite Fourier modes are no longer orthogonal. Various methods
have devised to restore orthogonality, such as the inclusion of a
cyclic prefix. \cite{vanNeeOFDM_book} These issues will be ignored
here and we will use the discrete Fourier mode version of
(\ref{eq:channel_eq_cont_FT}) given by
\begin{eqnarray}\label{eq:y_n_ft}
\hat{\vec y}_{pk} =  \bGhat_k \hat{\vec x}_{pk} + \hat{\vec
z}_{pk}
\end{eqnarray}
where the index $p$ represents the symbol index, $k$ is the
Fourier mode index with $k=0,\ldots,M-1$, $\bGhat_k$ is the
corresponding channel Fourier component for $\omega_k = 2\pi
k/(M\tau)$. $\hat{\vec x}_{pk}$ (and similarly $\hat{\vec
y}_{pk}$, $\hat{\vec z}_{pk}$) have been normalized so that
$\hat{\vec z}_k$, the Fourier transform of the noise vector $\vec
z_n$ is i.i.d. with elements $\sim {\cal CN}(0,1)$. Also, the
input signal in each frequency component $\hat{\vec x}_k$ is
assumed Gaussian with covariance $E\left[\hat{\vec
x}^{\phantom{\dagger}}_k \hat{\vec x}^\dagger_{k'}\right] =
\delta_{kk'} \vec Q_k$, normalized so that $\Tr{\vec Q_k} = \nt$.
For completeness, we rewrite the Fourier transform of the channel
in (\ref{eq:FourierTrans_channel}) as
\begin{eqnarray}\label{eq:G_k_ft_def}
  \bGhat_{k} = \sum_{l=0}^{L-1} \bG_l e^{i\frac{2\pi k m_l}{M}}
\end{eqnarray}
As mentioned earlier, the channel matrices $\bGhat_{k}$ are
assumed to be known at the receiver but not the transmitter.

\subsection{Channel Statistics}
\label{sec:Channel Statistics}

Next, we would like to characterize the statistics of the channel
matrices $\bG_l$ in (\ref{eq:basic_channel_eq}), which are random
due to fading. In particular, they are assumed to be zero-mean,
independent Gaussian random matrices. In addition, we assume the
correlations between elements of $\bG_l$ to be as follows:
\begin{equation}\label{eq:G_lG_l*_gen_corrs}
  \left<G^{\phantom{*}}_{l,i\alpha}G_{l',j\beta}^*\right>=\delta_{ll'}\,
  \frac{\rho_l}{\nt}\,
  T_{l,ij} \, R_{l,\alpha\beta}
\end{equation}
where the expectation $\left<\,\cdot\,\right>$ is over the fading
matrices $\bG_l$. $\rho_l$, $\vec T_l$ and $\vec R_l$ are the
signal to noise ratio, and  the $\nt$- and $\nr$-dimensional
correlation matrices for the $l$-th path at the transmitter and
receiver, respectively. Underlying the structure of the above
correlations is the assumption that different paths have
uncorrelated channels.\cite{Moustakas2001_WidebandMeasurements}
Each path is assumed to have correlations in the form of a
Kronecker product. This is certainly valid when each path
corresponds to a single scattered wave, in which case each of the
corresponding correlation matrices have unit rank. The above
channel model is in agreement with adopted channel models in third
generation standards \cite{SCM_3GPP_TR25_996}.

We comment that an interesting special case occurs when {\em all}
the delays $m_l$ take the same value, i.e. arrive within the same
$\tau$ interval (see section \ref{sec:Special Case 2: Narrowband
Multipath}). This represents a {\it narrowband} channel with
non-Kronecker product (or non-factorizable) correlations.  In
other words we could write the analogous simple narrowband channel
relation
\begin{equation}
\label{eq:narrowbandch1} \bf y_p = \bf G \bf x_p + \bf z_p
\end{equation}
where %
\begin{equation} \label{eq:narrowbandch2}{\bf G} = \sum_l
{\bf G}_l
\end{equation}
The matrices ${\bf G}_l$ have correlations of the form
(\ref{eq:G_lG_l*_gen_corrs}) above. We note that such a form
includes general models of polarization mixing with general
correlation matrices between the different polarization components
\cite{SCM_3GPP_TR25_996, Shafi2005_3DPolarizedMIMOChannels}. For
example, the correlations of a multipath channel with antennas of
different polarizations can be written compactly as
\begin{eqnarray}
\label{eq:gaussian_correlations_combined} %
\left<G_{i\alpha}G^*_{j\beta}\right> &=& \sum_{l}
\frac{\rho_l}{n_t}
\left[\begin{array}{c} 
T^{l,v}_{\alpha\beta} \\ 
T^{l,h}_{\alpha\beta}  
\end{array} \right]^T 
\left[\begin{array}{cc} 
1 & x \\ 
x & 1 
\end{array} \right] 
\left[\begin{array}{c} 
R^{l,v}_{ij} \\ 
R^{l,h}_{ij}  
\end{array} \right] 
\end{eqnarray}
where the sum is over all paths, $\rho_l$ is the signal-to-noise
ratio of each path $l$, $\vec T^{l,v}$, $\vec T^{l,h}$ are the
correlation matrices of the vertical and horizontal polarization
components of transmitter antennas for the $l$th path (and
similarly for the receiver arrays) and $x$ is the polarization
mixing ratio.

Finally, it should be stressed that the most general narrowband
zero-mean Gaussian model, including the recently proposed
independent non-identically distributed (IND) channel, can be
expressed in the form of (\ref{eq:G_lG_l*_gen_corrs}),
(\ref{eq:narrowbandch2}), since the correlations of any Gaussian
zero-mean matrix can be written as
\begin{equation}\label{eq:GG*_gen_narrowband_corrs}
  \left<G^{\phantom{*}}_{i\alpha}G_{j\beta}^*\right>=\sum_l \,
  T_{l,ij} \, R_{l,\alpha\beta}
\end{equation}
To see this, let $l=1,\ldots,(\nt\nr)^2$ and then set the matrices
$\vec T_l$, $\vec R_l$ have zero entries except for the element
$ij$ and $\alpha\beta$, respectively, when the index $l$ takes the
value
$l(i,j,\alpha,\beta)=i+\nt(j-1)+\nt^2(\alpha-1)+\nt^2\nr(\beta-1)$.
The non-zero values of these matrices can be chosen to be, for
example,
$T_{l(i,j,\alpha,\beta),ij}=\left<G^{\phantom{*}}_{i\alpha}G_{j\beta}^*\right>$
and $R_{l(i,j,\alpha,\beta),\alpha\beta} = 1$. Although this
mapping is not unique, it demonstrates the generality of our
method.

It should be noted that, since the receiver/mobile terminal is
usually assumed to be located deep inside the clutter, the
received signal tends to have wide angle-spread,  thereby making
the differences in the angles of arrival of different paths less
distinguishable. Therefore, it is sometimes reasonable to assume
that the receiver correlations $\vec R_l$ are path-independent,
i.e.
\begin{equation}\label{eq:G_lG_l*_corrs_R_l_indep}
  \left<G^{\phantom{*}}_{l,i\alpha}G_{l',j\beta}^*\right>=\delta_{ll'}\,
  \frac{\rho_l}{\nt}\,
  T_{l,ij} \, R_{\alpha\beta}
\end{equation}
This assumption is not as easily met at the transmitter/base
station, where the nearest scatterers are typically further
separated, thereby making the $\vec T_l$ typically different. A
further simplification of the above is the case when the receive
antennas are uncorrelated, which is discussed in
\cite{Bolcskei2002_OFDMMIMOCapacity}.

As a result of the above, $\bGhat_k$, the Fourier transform of
$\vec G_l$ (\ref{eq:G_k_ft_def}) is also Gaussian with
correlations
\begin{eqnarray}\label{eq:G_kG_k*_gen_corrs}
  \left<\hat{G}_{k,i\alpha}\hat{G}_{k',j\beta}^*\right>&=&
  \frac{1}{\nt}
  \sum_{l=0}^{L-1} \rho_l\, e^{i\frac{2\pi(k-k')m_l}{M}}\,
  T_{l,ij} R_{l,\alpha\beta}
\end{eqnarray}

For the case of narrowband channels mentioned above in
(\ref{eq:narrowbandch1}), (\ref{eq:narrowbandch2}), $\bGhat_k$ is
nonzero only for $k=0$, therefore $\bGhat_0 = \bf G$ with $\bG$
given in (\ref{eq:narrowbandch2}).

\section{Wideband Mutual Information}
\label{sec:Mutual Information}

The mutual information of each of the frequency components $k$ is
given by\cite{Foschini1998_BLAST1, Telatar1995_BLAST1}
\begin{eqnarray}\label{eq:mut_info_nb_def}
I_k =   \log\det\left(\vec I_\nr + \bGhat_k \vec Q_k
\bGhat_k{}^\dagger \right)
\end{eqnarray}
The $\log$ above (and throughout the whole paper) represents the
natural logarithm and thus $I$ is expressed in nats. The total
mutual information over all frequency components is then
\begin{eqnarray}\label{eq:mut_info_wb_def}
I &=& \sum_{k=0}^{M-1} I_k
\end{eqnarray}

\subsection{Statistics of Mutual Information}
\label{sec:Statistics of Mutual Information}

The distribution of the mutual information can be characterized
through its moments. These moments can be evaluated by first
calculating the moment generating function $g(\nu)$ of $I$
\begin{eqnarray}\label{eq:g_nu_def}
  g(\nu)  &=& \left< \left[\prod_{k=0}^{M-1}
  \det \left(\vec I_\nr + \bGhat_k \vec Q_k \bGhat_k{}^\dagger\right)
  \right]^{-\nu}\right> \\ \nonumber
  &=& \left<  e^{-\nu I}  \right> %
  \\ \label{eq:g_nu_def_expanded} %
    &=& 1 - \nu \left< I \right> + \frac{\nu^2}{2} \left< I^2 \right> + \ldots
\end{eqnarray}
Assuming that $g(\nu)$ is analytic at least in the vicinity of
$\nu=0$, we can express $\log g(\nu)$ as follows
\begin{equation} \label{eq:logg_nuexpansion}
\log g(\nu) =\sum_{p=1}^\infty \frac{(-\nu)^p}{p!} {\cal C}_p
\end{equation}
where ${\cal C}_p$ is the $p$-th cumulant moment of $I$.   For
example, the ergodic mutual information, i.e. the average of the
distribution is given by
\begin{eqnarray}
{\cal C}_1 &=& \left< I \right>  = \sum_{k=0}^{M-1} \langle I_k
\rangle \\ &=& \label{eq:line22}
 \sum_{k=0}^{M-1} \left<\log\det\left(\vec I_\nr + \bGhat_k
\vec Q_k \bGhat_k{}^\dagger \right)\right>
\label{eq:ave_mut_info_wb_def}
\end{eqnarray}
Similarly, the variance of the distribution is
\begin{eqnarray}
\label{eq:C2def} {\cal C}_2 &=&  {Var(I)} = \left< (I - \langle I
\rangle)^2 \right>
\\ &=&  \left(\sum_{k,k'=0}^{M-1} \langle I_k I_{k'}\rangle \right)
- \left< I \right>^2 \label{eq:kkp}
\end{eqnarray}
its the skewness of the distribution is
\begin{eqnarray}
{\cal C}_3 = Sk(I)=\langle\left(I-\langle I \rangle
\right)^3\rangle
\end{eqnarray}
and so forth. Note that since $I_k$ depends only on $\bGhat_k$ and
$\vec Q_k$, to evaluate the ergodic average
(\ref{eq:ave_mut_info_wb_def}) we can perform the average for each
term in the sum in (\ref{eq:line22}) separately, neglecting any
correlations between $\bGhat_k$'s with different $k$ indices. Thus
for evaluating the ergodic average, the only correlation of
relevance is $\langle\hat{G}_{k,i\alpha}\hat{G}_{k,j\beta}\rangle$
which turns out to be $k$-independent, as seen in
(\ref{eq:G_kG_k*_gen_corrs}). Therefore the only $k$-dependence of
each term in the sum in (\ref{eq:ave_mut_info_wb_def}) is through
$\vec Q_k$. As a result the optimal $\vec Q_k$ will be
$k$-independent. We will thus henceforth assume that $\vec Q_k$ is
chosen to be a $k$-independent quantity $\vec Q$. As a result, the
wideband ergodic capacity becomes just $M$ times its narrowband
counterpart \cite{Bolcskei2002_OFDMMIMOCapacity}. This
$k$-independence of the mean mutual information will be of use in
the next section. In contrast, in evaluating higher moments of the
distribution such as the variance, as is easily seen from
(\ref{eq:kkp}), we will have to consider cross correlations
between $\bGhat_k$ and $\bGhat_{k'}$

Finally,it should be emphasized that the distribution of the
mutual information can also be completely characterized by the
outage mutual information\cite{Ozarow1994_OutageCapacity},
obtained by inverting the expression below with respect to
$I_{out}$
\begin{equation}\label{eq:outage_mut_info_def}
  P_{out}=\mbox{Prob}\left(I<I_{out}\right)
\end{equation}
where $Prob(I<I_{out})$ is the probability that the mutual
information is less than a given value $I_{out}$.

\section{Mathematical Framework}
\label{sec:math_framework}

The purpose of this paper is to analyze the statistics of the
wideband mutual information $I$ in (\ref{eq:mut_info_wb_def}) for
general zero-mean Gaussian channels. In this section we describe
the basic steps to derive analytic expressions for the first few
cumulant moments of $I$, valid formally for large antenna numbers.
In this limit it has been shown elsewhere
\cite{Moustakas2003_MIMO1,
Hochwald2002_MultiAntennaChannelHardening,
Wang2002_OutageMutualInfoOfSTMIMOChannels,
Smith2002_OnTheGaussianApproximationToTheCapacityOfWirelessMIMOSystems}
that the narrowband mutual information distribution becomes
asymptotically Gaussian. Thus the first two moments can describe
the outage mutual information (\ref{eq:outage_mut_info_def}).
Using the mathematical framework of \cite{Moustakas2003_MIMO1,
Sengupta2000_BLAST1} we will show that this Gaussian character
holds also for wideband channels.

To obtain the moments of the mutual information distribution we
need to calculate $g(\nu)$ in (\ref{eq:g_nu_def}) for $\nu$ in the
vicinity of $\nu=0$. To achieve this we will employ the replica
assumption discussed in \cite{Sengupta2000_BLAST1,
Sengupta1999_BLAST1, Moustakas2000_BLAST1, Moustakas2003_MIMO1}.
\begin{assumption}[Replica Method]
\label{ass:replica} $g(\nu)$ evaluated for positive integer values
of $\nu$ can be analytically continued for real $\nu$,
specifically in the vicinity of $\nu=0^+$.
\end{assumption}

This assumption, used also in
\cite{ParisiBook,Itzykson_book_StatisticalFieldTheory,
Muller2002_RandomMatrixMIMO,
Guo2003_ReplicaAnalysisOfLargeCDMASystems,
Tanaka2002_ReplicasInCDMAMUD}, alleviates the problem of dealing
with averages of logarithms of random quantities, since the
logarithm is obtained after calculating $g(\nu)$.

In Appendix \ref{app:Derivation of gnu_S} we show that $g(\nu)$
can be expressed as an integral over $M\nu\times M\nu$ complex
matrices $\bRcal^l$, $\bTcal^l$, with $l=0,\ldots,L-1$
\begin{equation} \label{eq:gnu_S}
   g(\nu)  =
  \int d\mu\left(\{ \bTcal^l ,\bRcal^l \} \right)
     e^{-{\cal S }}
\end{equation}
where the integration metric was defined in (\ref{eq:dmuTR1}) and
\begin{eqnarray}
 {\cal S} &=& \log\det \left(\vec I_\nr \outer \vec I_{\nu M} +
  \sum_l \frac{1}{\sqrt{\nt}} {\bf R}_l \outer \bRcal^l \right)
\nonumber \\
 &+& \log\det \left(\vec I_\nr \outer \vec I_{\nu M} +  \sum_l\frac{\rho_l}{\sqrt{\nt}}
{\vec Q} {\bf T}_l \outer \hat \bTcal^l \right)
\nonumber \\%
 &-& \sum_l \Tr{\bTcal^l \, \bRcal^l}
\label{eq:S_full}
\end{eqnarray}
where $\hat \bTcal^l$ is an $M \nu \times M \nu$ matrix related to
$\bTcal^l$ via
\begin{equation}
\hat \bTcal^l_{k \alpha; k'\beta} = \bTcal^l_{k \alpha; k' \beta}
e^{\frac{2\pi i (k-k')m_l}{M}}
\end{equation}
where we have explicitly written out the components of the
matrices here with $k,k'$ ranging from $0$ to $M-1$ and  $\alpha$
and $\beta$ ranging from $1$ to $\nu$.  (See the notation in
Appendix \ref{app:Derivation of gnu_S}).

At this point $\nu$ is still a positive integer, which has to be
taken to zero following Assumption \ref{ass:replica}, in order to
be able to expand $g(\nu)$ for small $\nu$, as in
(\ref{eq:g_nu_def_expanded}). However, since the integral in
(\ref{eq:gnu_S}) cannot be performed exactly, we need to calculate
it asymptotically in the limit of large antenna numbers
$\nt$,$\nr\gg1$. Therefore we need to interchange the limits
$n\gg1$ and $\nu\rightarrow 0^+$.
\begin{assumption}[Interchanging Limits]\cite{Moustakas2003_MIMO1}
\label{ass:interchanging_limits} The limits $n\rightarrow \infty$
and $\nu\rightarrow 0^+$ in evaluating $g(\nu)$ in
(\ref{eq:gnu_S}) can be interchanged by first taking the former
and then the latter without affecting the final answer.
\end{assumption}

\subsection{Saddle-Point Analysis}
\label{sec:saddle_pt_soln}

We now use Assumption \ref{ass:interchanging_limits} to calculate
(\ref{eq:gnu_S}) asymptotically for large $\nt$, $\nr$, by
deforming the integrals in (\ref{eq:gnu_S}) to pass through a
saddle point. More details are given in
\cite{Moustakas2003_MIMO1}. To specify the structure of the
saddle-point solution, i.e. the form of $\bTcal^l$, $\bRcal^l$ at
the saddle-point, we assume as in \cite{Moustakas2003_MIMO1} that
the relevant saddle-point solution is invariant in
$\nu$-dimensional replica space. However, in our case since
$\bTcal^l$, $\bRcal^l$ are $\nu M$-dimensional matrices, this is
not enough to fully characterize the saddle-point. Therefore, we
will also assume that the saddle-point values of $\bTcal^l$,
$\bRcal^l$ are invariant in $M$-dimensional frequency $q$-space.
This assumption, as we shall see, leads to a saddle-point value of
${\cal S}$, and to an ergodic average of the mutual information,
that is independent of inter-frequency correlations, in agreement
with the correct answer, as discussed in Section
\ref{sec:Statistics of Mutual Information} and
\cite{Bolcskei2002_OFDMMIMOCapacity}.

Thus, at the saddle-point $\bTcal^l$, $\bRcal^l$ take the form
$\bTcal^l = t_l \sqrt{\nt} \, \vec I_{\nu M}$, $\bRcal^l= r_l
\sqrt{\nt}\, \vec I_{\nu M}$, where $t_l$ and $r_l$ are positive,
still undetermined numbers of order unity in the number of
antennas. A scaling factor of $\sqrt{\nt}$ has been included for
convenience, as will become evident below. Following
\cite{Moustakas2003_MIMO1} we analyze the integral in
(\ref{eq:gnu_S}) by shifting the origin of integration to the
saddle point, i.e. by rewriting $\bTcal$, $\bRcal$ as
\begin{eqnarray}
\label{eq:TR_saddle_point_vicinity} %
\bTcal^l &=& t_l \sqrt{\nt} \, \vec I_{\nu M} + \dTcal^l \\
\nonumber %
\bRcal^l &=& r_l  \sqrt{\nt} \, \vec I_{\nu M} + \dRcal^l
\end{eqnarray}
where $\dTcal^l$, $\dRcal^l$ are $\nu M$-dimensional matrices
representing deviations around the saddle point. One can then
expand ${\cal S}$ in (\ref{eq:S_full}) in a Taylor series of
increasing powers of $\dTcal^l$, $\dRcal^l$ as follows
\begin{equation}
\label{eq:expansion_S}%
{\cal S} = {\cal S}_0 + {\cal S}_1 + {\cal S}_2 + {\cal S}_3
\ldots
\end{equation}
with ${\cal S}_p$ containing $p$-th order terms in $\dTcal^l$,
$\dRcal^l$. These terms are shown explicitly in Appendix
\ref{app:Details for Saddle Point Analysis of gnu_S} in
(\ref{eq:saddle0}), (\ref{eq:S1_expanded}),
(\ref{eq:S2_expanded}), (\ref{eq:S_p_greater2}), where it can be
seen that ${\cal S}_p$ is ${\cal O}(n^{1-p/2})$, making
(\ref{eq:expansion_S}) indeed an asymptotic expansion in inverse
powers of $n$.

The saddle point solution of  (\ref{eq:gnu_S}) and hence the
corresponding values of $t_l$, $r_l$ is found by demanding that
${\cal S}$ is stationary with respect to variations in $\bTcal^l$,
$\bRcal^l$. \cite{Bender_Orszag_book} This means that ${\cal S}_1
= 0$ (see (\ref{eq:S1_expanded})), which is analogous to setting
the first derivative of a function to zero, in order to find its
maximum or minimum. This produces the following saddle-point
equations:
\begin{eqnarray}
\label{eq:r_gen_case} %
 r_l &=& \frac{\rho_l}{\nt} \Tr{ \vec Q \vec T_l \left[\vec I_{\nt} +
 \vec Q {\tilde {\bf T}} \right]^{-1}}
  \\ \label{eq:t_gen_case} %
  t_l &=& \frac{1}{\nt} \Tr{\vec R_l \left[\vec I_\nr+{\tilde {\bf R}} \right]^{ -1}  }
\end{eqnarray}
where ${\tilde {\bf T}}$, ${\tilde {\bf R}}$ have been defined as
\begin{eqnarray}
\label{eq:T_tilde} %
 {\tilde {\bf T}} &=& \sum_{l}\rho_{l} t_{l}\vec T_{l}
  \\ \label{eq:R_tilde} %
  {\tilde {\bf R}} &=& \sum_{l} r_{l}\, \vec R_{l}
\end{eqnarray}

The next term in the expansion of ${\cal S}$ is ${\cal S}_2$ and
needs to be taken into account non-perturbatively, because it is
${\cal O}(1)$ in the number of antennas $n$ and thus will provide
a finite correction.  Fortunately, ${\cal S}_2$ is quadratic in
the variables $\dTcal^l$ and $\dRcal^l$ so that the integral
(\ref{eq:gnu_S}) is just a Gaussian integral at this order.

In contrast, ${\cal S}_p$ terms with $p>2$ become vanishingly
small at large $n$, since they are ${\cal O}(n^{1-p/2})$.
Therefore, they can be expanded from the exponent in
(\ref{eq:gnu_S}) and treated perturbatively as follows:
\begin{eqnarray}
\label{eq:g_nu_S_expansion} %
g(\nu)  &=& e^{-{\cal S}_0} \int d \mu\left(\{\dTcal^l\},
\{\dRcal^l\} \right) e^{-{\cal S}_2} \\ \nonumber %
&&\cdot \left(1
- {\cal S}_3 - {\cal S}_4 +\frac{1}{2} {\cal S}_3^2 + \ldots
\right)
\end{eqnarray}
Each term in this expansion can be evaluated explicitly, with
higher order terms producing corrections of increasingly higher
orders in $1/n$. Subsequently, taking the logarithm of the result
as prescribed in (\ref{eq:logg_nuexpansion}) will produce an
$1/n$-expansion for the cumulant moments of $I$, with only integer
powers of $1/n$ surviving \cite{Moustakas2003_MIMO1}).

\subsection{Ergodic Capacity}
\label{sec:ergodic_capacity}

{}From  (\ref{eq:saddle0}) in Appendix \ref{app:Details for Saddle
Point Analysis of gnu_S} we see that ${\cal S}_0=\nu\Gamma$ with
proportionality factor $\Gamma$ being the leading term to the
mutual information, which is given by
\begin{eqnarray}
\label{eq:ergodic_mutual_information_1storder}
 \Gamma &=&
     M \log \det \left( \vec I_\nt + \,
      \vec Q {\tilde {\bf T}}   \right)
        \\ \nonumber
       &+& M\log \det \left(\vec I_\nr + {\tilde {\bf R}}  \right)  - \nt M \sum_l r_l t_l
\end{eqnarray}
where $t_l$, $r_l$, ${\tilde {\bf T}}$, ${\tilde {\bf R}}$ are
given by (\ref{eq:t_gen_case}), (\ref{eq:r_gen_case}),
(\ref{eq:T_tilde}), (\ref{eq:R_tilde}).

Note that the above equations are independent of the relative
delays between paths, thereby applying to narrowband channels, as
well as wideband channels with non-trivial delays between paths.
This is to be expected since the ergodic wideband capacity is
independent of delay. \cite{Bolcskei2002_OFDMMIMOCapacity}

To obtain the capacity-achieving input distribution $\vec Q$,
$\langle I\rangle$ has to be optimized subject to the power
constraint $\mbox{Tr}\{\vec Q\} = \nt$.  This constraint is
enforced by adding a Lagrange multiplier to $\langle I\rangle$,
i.e.
\begin{eqnarray}
\label{eq:lagrange_multiplier}
 \langle I\rangle &\rightarrow& \langle I\rangle
- \Lambda \left( \Tr{\vec Q} -\nt\right) \\ \nonumber %
&=&\langle I\rangle - \Lambda \left( \sum_i q_i -\nt\right)
\end{eqnarray}
where $q_i$ are the $\nt$ eigenvalues of $\vec Q$.   As in
\cite{Moustakas2003_MIMO1}, the eigenvectors of the optimal $\vec
Q$ are the same as ${\tilde {\bf T}}$ (at least to ${\cal
O}(1/n)$). This statement is proven in Appendix
\ref{app:optimalQ}.

With the constraint that $\vec Q$ and $\tilde {\bf T}$ should be
diagonal in the same basis, we can find the optimal $\vec Q$ by
differentiating with respect to the eigenvalues $q_i$. It is then
easy to see \cite{Moustakas2003_MIMO1} that the optimal
eigenvalues of $\vec Q$ are given by
\begin{equation}
\label{eq:qeigs}
        q_i = \left[ \frac{1}{\Lambda}  - \frac{1}{{\tilde T}_i} \right]_+
\end{equation}
where ${\tilde T}_i$ are the $\nt$ eigenvalues of ${\tilde {\bf
T}}$ and $\left[ x\right]_+ = \{x + \mbox{sgn}(x) \}/2$. Here, the
Lagrange multiplier $\Lambda>0$ is determined by imposing the
power constraint
\begin{equation}
    \label{trsum}
    \Tr{\vec Q} = \sum_{i=1}^{\nt} q_i = \nt
\end{equation}
with $q_i$ given by (\ref{eq:qeigs}).

\subsection{Variance of the Mutual Information}
\label{sec:variance_mutual_info_iid}

To obtain the ${\cal O}(\nu^2)$ term in the expansion of $\log
g(\nu)$ in  (\ref{eq:logg_nuexpansion}) we need to only include
the next non-vanishing term, ${\cal S}_2$. The second line in
(\ref{eq:g_nu_S_expansion}) can be temporarily neglected.

 Using
the saddle point value for ${\cal S}_0 = \nu \Gamma$ in
Eq.(\ref{eq:g_nu_S_expansion}), the integration over $\dRcal^l$,
$\dTcal^l$ can be performed straightforwardly (see
\cite{Moustakas2003_MIMO1} for more details), resulting in
\begin{eqnarray}
\label{eq:g_nu_S0_S2} %
g(\nu) &=& e^{-\nu \Gamma} \prod_{k,k'}\left|\det \vec
V^{kk'}\right|^{-\frac{\nu^2}{2}}
\end{eqnarray}
where the $2L$-dimensional matrix  $\vec V^{kk'}$ is given in
Appendix \ref{app:Details for Saddle Point Analysis of gnu_S} by
(\ref{eq:hessian}). Thus, by comparing (\ref{eq:logg_nuexpansion})
to (\ref{eq:g_nu_S0_S2}) and by matching order by order the terms
of the $\nu$-Taylor expansion of $\log g(\nu)$, the leading term
in the variance of the mutual information is
\begin{eqnarray}
\label{eq:variance_leading_order}
   {\cal C}_2 &=& Var(I) = -\sum_{kk'} \log|\det\vec V^{kk'}|  + {\cal O}(1/ n^{2}) \\ \nonumber%
    &=& -\sum_{kk'}\log\det\left(\vec I_L - \vec M_{r,2}^{1/2} \vec M_{t,2} \vec
    M_{r,2}^{1/2}\right)+ {\cal O}(1/ n^{2})
\end{eqnarray}
where the $L$-dimensional matrices $\vec M_{t,2}$, $\vec M_{r,2}$
are given in Appendix \ref{app:Details for Saddle Point Analysis
of gnu_S} by (\ref{eq:Mdef}) and (\ref{eq:Mdef2}). We note that
since $\vec M_{r,2}$ and  $\vec M_{t,2}$ are both ${\cal O}(1)$,
the variance is also formally ${\cal O}(1)$ in the $1/n$ expansion
when both $\nt$ and $\nr$ are of the same order.

\subsection{Higher Order Terms} \label{sec:higher_order_terms}

To obtain higher-order corrections in $n^{-1}$, beyond the ${\cal
O}(n)$ and ${\cal O}(1)$ terms that appear in the average and the
variance, respectively, one needs to take into account the terms
${\cal S}_p$ with $p>2$ in (\ref{eq:g_nu_S_expansion}). These
terms will give rise to higher-order cumulant moments of the
distribution of the mutual information, as well as higher-order
corrections to the first two cumulant moments. In Appendix
\ref{app:higher_order_terms} we sketch the calculation of the next
leading correction terms of order ${\cal O}(n^{-1})$. Including
this additional term $g(\nu)$ can be written as
\begin{equation}
  g(\nu) =  e^{-\nu \Gamma} \prod_{k,k'}\left| \det {\bf
    V}^{kk'} \right|^{-\nu^2/2} \left[ 1 + D_1 + {\cal
    O}(n^{-2})\right] \label{eq:Dexpansion}
\end{equation}
where $D_1$ is given by
\begin{equation}
    \label{eq:D_1}
    D_1 = a_1 \nu + a_3 \nu^3
\end{equation}
and $a_1$ and $a_3$ are defined in (\ref{eq:a1def}),
(\ref{eq:a3def}), which are indeed ${\cal O}(1/n)$.

Using the cumulant expansion notation of
(\ref{eq:logg_nuexpansion}) and matching the generated terms above
to the appropriate powers of $\nu$, we see that $D_1$ produces
order ${\cal O}(1/n)$ terms to the first cumulant (mean) ${\cal
C}_1$ and third cumulant (skewness) ${\cal C}_3$:
\begin{eqnarray}\label{eq:mean_correction}
 {\cal C}_1 &=& \Gamma - a_1 + {\cal O}(1/n^3)
\\ \label{eq:skewness_term}
 {\cal C}_3 &=& -6a_3 + {\cal O}(1/n^3)
\end{eqnarray}

\subsection{Special Case 1: $\vec R_l$ independent of $l$}
\label{sec:special case: R_l independent of l}

In this section, we will show how the above results simplify when
the correlation matrix of the receiver or  transmitter is
independent of the path index $l$. For concreteness we will only
analyze the case where $\vec R_l$ is independent of $l$, i.e. when
the channel correlations take the form
(\ref{eq:G_lG_l*_corrs_R_l_indep}).

In this case we see that in (\ref{eq:t_gen_case}) $t_l$ is
independent of the path index $l$ and thus we may set $t=t_l$.
Furthermore, by summing (\ref{eq:r_gen_case}) over $l$ we get
\begin{eqnarray}
\label{eq:r_R_indep_l} %
 r \equiv \sum_l r_l &=& \frac{1}{\nt} \Tr{ \frac{\vec Q \vec T}{\vec I_{\nt} +
 t \vec Q \vec T} }
  \\ \label{eq:t_R_indep_l} %
  t &=& \frac{1}{\nt} \Tr{\frac{\vec R}{ \vec I_\nr+ r {\bf R}}}
\end{eqnarray}
where ${\bf T} = \sum_l \rho_l \vec T_l$ and $\vec R=\vec R_l$.
Thus the mutual information in
(\ref{eq:ergodic_mutual_information_1storder}) may be written as
\begin{eqnarray}
\label{eq:ergodic_mutual_information_1storder_R_indep l}
 \left<I\right> &=&
     M \log \det \left( \vec I_\nt + \,
      t \vec Q  {\bf T}   \right)
        \\ \nonumber
       &+& M\log \det \left(\vec I_\nr + r {\bf R}  \right)  - M \nt r t
\end{eqnarray}
Note that, apart from a redefinition of $\vec T$ to take into
account multiple paths, these results are identical to those
derived previously for narrowband channels
\cite{Moustakas2003_MIMO1}.

To derive a simplified expression for the variance from
(\ref{eq:variance_leading_order}), we note that $M_{r,2}$ now
becomes a constant matrix, which can be written as a vector outer
product
\begin{eqnarray}\label{eq:M_r2_outer_prod_form}
\vec M_{r,2}^{ll'} = \frac{1}{\nt} \Tr{\left(\frac{\vec R}{\vec
I_\nr + r \vec R}\right)^2} \vec v \vec v^\dagger= m_{r,2}  \vec v
\vec v^\dagger
\end{eqnarray}
where the vector $\vec v$ has elements $v_l =1 $ for all
$l=1,\ldots,L$. The second equality in the above equation defines
$m_{r,2}$. Similarly, $M_{t,2}$ can be written as:
\begin{eqnarray}\label{eq:M_t2_outer_prod_form}
M_{t,2}^{ll'} &=& \frac{\rho_l\rho_{l'}}{\nt}
\exp\left[\frac{2\pi i
(k_1-k_2)(m_l-m_{l'})}{M}\right] \\ \nonumber 
&\cdot& \Tr{\frac{1}{\vec I_\nt + t \vec Q \vec T} \vec Q\vec T_l
\frac{1}{\vec I_\nt + t \vec Q \vec T} \vec Q\vec T_{l'}}
\end{eqnarray}
After some algebra we see that (\ref{eq:variance_leading_order})
simplifies to
\begin{eqnarray}
\label{eq:variance_R_indep_l}
   Var(I) &=& -\sum_{kk'} \log\left|1 - m_{r,2} \, m_{t,2}^{k-k'}\right|
\end{eqnarray}
where
\begin{eqnarray}\label{eq:m_t2_def}
m_{t,2}^{q} = \frac{1}{\nt} \Tr{\frac{1}{\vec I_\nt + t \vec Q
\vec T} \vec Q\vec S_q \frac{1}{\vec I_\nt + t \vec Q \vec T} \vec
Q\vec S_{-q}}
\end{eqnarray}
with the matrix $\vec S_q$ defined as
\begin{eqnarray}\label{eq:m_S_q_def}
\vec S_{q} = \sum_l \rho_l \vec T_l \exp\left(\frac{2\pi i q
m_l}{M}\right)
\end{eqnarray}
which is the temporal Fourier transform of the correlation
matrices $\vec T_l$.

\subsection{Special Case 2: Narrowband Multipath}
\label{sec:Special Case 2: Narrowband Multipath}

As mentioned in the introduction, this approach is applicable in
calculating the ergodic average and variance of an arbitrary
Gaussian zero-mean channel. This obviously includes a narrowband
channel with arbitrary correlations. The only difference in the
analysis of this channel is that all delay indices $m_l$ are equal
and can thus be set to zero.

\section{Analysis of Results} \label{sec:Analysis of Results}

In the previous section we have seen that in the limit of large
antenna numbers $n$, the mean mutual information is of order $n$,
while the variance of the distribution is of order unity. In
addition, in Appendix \ref{app:higher_order_terms} we find that
the skewness (the third cumulant moment) is ${\cal O}(1/n)$ and
higher cumulant moments are even smaller (${\cal O}(1/n^2)$). In
agreement with the narrowband case
\cite{Hochwald2002_MultiAntennaChannelHardening,
Moustakas2003_MIMO1}, this suggests that the distribution of the
wideband multipath mutual information is also Gaussian for large
$n$. This Gaussian behavior was seen to be very accurate even for
small antenna arrays for narrowband channels
\cite{Moustakas2003_MIMO1,
Smith2002_OnTheGaussianApproximationToTheCapacityOfWirelessMIMOSystems}.
Below, we will see this to hold also in wideband multipath
channels by numerically comparing  the Gaussian distribution
${\cal N}\left[\langle I\rangle, Var(I)\right]$ calculated using
(\ref{eq:saddle0}) and (\ref{eq:variance_leading_order}) with the
simulated distribution resulting from the generation of a large
number of random matrix realizations. We will specifically analyze
four representative situations to show the effects of multipath on
the distribution of the mutual information of wideband channels.

If the distribution of the mutual information is Gaussian, we can
express $I_{out}$ from (\ref{eq:outage_mut_info_def}) as
\begin{equation}\label{eq:outage_mut_info_gaussian_def}
  I_{out} = \left<I\right> - \sqrt{2Var(I)}\Phi^{-1}(2P_{out}-1)
\end{equation}
where $\Phi^{-1}(x)$ is the inverse error
function.\cite{Abramowitz_Stegun_book} Clearly, this can only be
an approximation, since the mutual information cannot take
negative values.

\subsection{Distribution of Wideband Mutual Information for $L$ equal-power equally-spaced i.i.d. paths}
\label{sec:open_loop_iid}

It is instructive to apply the above results to the case of $L$
equal power paths, with $\rho_l=\rho/L$ in
(\ref{eq:G_kG_k*_gen_corrs}), with $\nt=\nr=n$ and with
correlation matrices being unity, i.e. $\vec R_l=\vec T_l=\vec
I_n$. Also, for simplicity we assume the delays of the paths are
all equally spaced from each other by $\tau$, i.e. $m_l=l$. This
is a special case of the one discussed in Section \ref{sec:special
case: R_l independent of l}. In this case the optimal input
distribution is $\vec Q=\vec I_n$ \cite{Moustakas2003_MIMO1}, and
(\ref{eq:ergodic_mutual_information_1storder_R_indep l}) becomes
\begin{equation}
\label{eq:I_ave_iid} \langle I \rangle = nM\left[ \log\left(1
+\rho t \right) + \log\left(1+r\right) -t r\right]
\end{equation}
with the extremizing values of $r$ and $t$ from
(\ref{eq:r_R_indep_l}), (\ref{eq:t_R_indep_l}) given by
\begin{eqnarray}
\label{eq:t_iid} r = \rho t = \frac{\sqrt{1+4\rho}-1}{2}
\end{eqnarray}
which gives
\begin{equation} \label{eq:I_ave_iid_optimal}
\langle I \rangle = nM\left[
2\log\left(\frac{\sqrt{1+4\rho}+1}{2}\right) -
\frac{\left(\sqrt{1+4\rho}-1\right)^2}{4\rho}\right]
\end{equation}
This result is identical to the one derived elsewhere
\cite{Moustakas2003_MIMO1,Verdu1999_MIMO1}. The variance can be
calculated using (\ref{eq:variance_R_indep_l}),
(\ref{eq:m_t2_def}) with the $\vec S_q$ in (\ref{eq:m_S_q_def})
taking the form $S_q=\rho\vec I_n/L$ and takes the form
\begin{equation}\label{eq:var_iid}
  Var(I) = -\sum_{k,k'=0}^{M-1}
  \log\left[1-\left(\frac{t\rho}{t\rho+1}\frac{\sin\frac{\pi L
  (k-k')q}{M}}{L\sin\frac{\pi(k-k')q}{M}}\right)^2\right]
\end{equation}
with $t$ given by (\ref{eq:t_iid}). We see that the larger $L$ is,
the more peaked the ratio inside the logarithm is, and therefore
the smaller the variance. If $L = M$, the ratio of sines in
(\ref{eq:variance_leading_order}) becomes proportional to a
Kronecker delta function $\delta_{kk'}$, so that the variance
becomes equal to
\begin{equation}\label{eq:var_iid_L=M}
  Var(I) = -M
  \log\left[1-\left(\frac{\sqrt{1+4\rho}-1}{\sqrt{1+4\rho}+1}\right)^2\right]
\end{equation}
In general we can say that the variance of the {\em normalized}
mutual information per channel (i.e. $I/M$) scales as
$Var(I/M)\sim 1/L$.

\subsection{Distribution of Wideband Mutual Information for an
exponentially distributed power delay profile}
\label{sec:open_loop_exponential_distributed_delay}

\begin{figure}[htb]
\centerline{\epsfxsize=1.0\columnwidth\epsffile{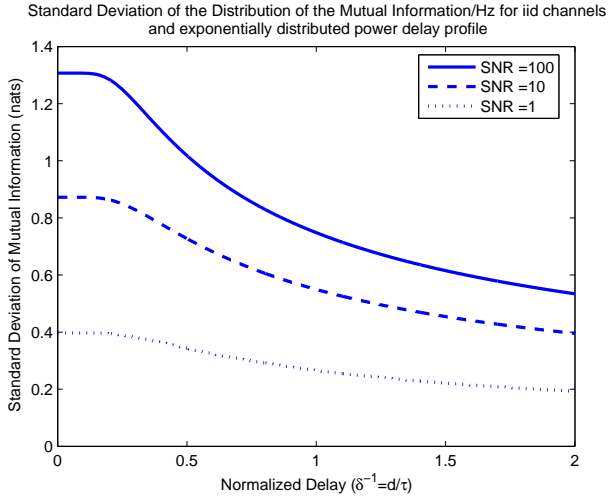}}
\caption{Standard deviation of the mutual information as a
function of the normalized delay spread ($d/\tau$) for the case of
an exponential power delay profile for three different
signal-to-noise ratios. For zero delay ($d=0$), the narrowband
result is recovered ($y$-axis). For increasing delays compared to
bandwidth $d>\tau$, the standard deviation of the mutual
information decreases. Eq. (\ref{eq:var_cont_exponential_profile})
has been used.} \label{fig:var_cap_expon_channel}
\end{figure}

We can also apply this approach to a more realistic version of a
multipath channel, namely one with an exponential power delay
profile, which can be expressed as
\begin{equation}\label{eq:rho_l_exponential_profile}
\rho_l = \bar{\rho} \left(1-e^{-\delta}\right)\,e^{-\delta l}
\end{equation}
where $\delta^{-1}=d/\tau$ is the product of the delay constant
$d$ with the bandwidth $\tau^{-1}$, and $\bar{\rho}$ is the
signal-to-noise ratio for the total power-delay profile. We have
implicitly assumed here that the number of paths is infinite,
$L=\infty$. For the simple case of uncorrelated channels, where
both $\vec T^l$ and $\vec R^l$ are unit matrices, the average
mutual information is identical to (\ref{eq:I_ave_iid}), by
replacing $\rho$ with $\bar{\rho}$. This can easily be seen  by
observing that the average mutual information in
(\ref{eq:ergodic_mutual_information_1storder_R_indep l}) is a
function of $\rho_l$ only through $\vec T$, which here is equal to
\begin{eqnarray}\label{eq:T_vec_iid_rho_bar}
\vec T = \sum_l\rho_l\vec T_l = \vec I_\nt \bar{\rho}
\end{eqnarray}
To calculate the variance of $I$, we first need to calculate
$m_{t,2}^{k-k'}$ and $m_{r,2}$ in (\ref{eq:variance_R_indep_l}).
The former can be evaluated from (\ref{eq:m_t2_def}) by performing
the sum (\ref{eq:m_S_q_def})
\begin{eqnarray}\label{eq:m_S_q_iid}
\vec S_{q} &=& \sum_{l=0}^\infty \rho_l \vec T_l e^{\frac{2\pi i q
m_l}{M}} = \vec I_\nt \sum_{l=0}^\infty \rho_l  e^{\frac{2\pi i q
m_l}{M}} \\ \nonumber %
&=& \bar{\rho} \frac{1-e^{-\delta}}{1-e^{-\delta+\frac{2\pi
qi}{M}}}
\end{eqnarray}
As a result $m^q_{t,2}$ (and similarly $m_{r,2}$) can be expressed
as
\begin{eqnarray}
\label{eq:mt2_exponential_profile} %
m_{t,2}^q &=& \frac{\bar{\rho}^2}{\left(1+t\bar{\rho}\right)^2}
\left|\frac{1-e^{-\delta}}{1-e^{-\delta+2\pi i
q/M}}\right|^2 \\
m_{r,2} &=& \frac{1}{\left(1+r\right)^2} = t^2
\end{eqnarray}
so that the normalized variance per channel can be expressed as
\begin{eqnarray}
\label{eq:var_discrete_exponential_profile} %
Var\left(\frac{I}{M}\right)  &=& - \frac{1}{M^2}\sum_{k,k'}
\log\left|1- m_{r,2} m_{t,2}^{k-k'}\right|
\end{eqnarray}
When the number of frequency channels $M$ is large, we can
approximate the above sums with integrals over frequency, which
can be performed analytically to give
\begin{eqnarray}
\label{eq:var_cont_exponential_profile} %
Var\left(\frac{I}{M}\right) &=&
-\ln\left\{\frac{1}{2}\left[1+e^{-2\delta} -
\beta\left(1-e^{-\delta}\right)^2 \right.\right. \\ \nonumber
&+&\left.\left.\sqrt{\left(1+e^{-2\delta}-\beta\left(1-e^{-\delta}\right)^2\right)-4e^{-2\delta}}\right]\right\}
\end{eqnarray}
where
\begin{eqnarray}
\label{eq:var_cont_exponential_profile_alpha_beta} %
 \beta &=& \left(\frac{t\bar{\rho}}{1+t\bar{\rho}}\right)^2
 = \frac{16\bar{\rho}^2}{\left(1+\sqrt{1+4\bar{\rho}}\right)^4}
\end{eqnarray}
(\ref{eq:var_cont_exponential_profile}) is plotted in Fig.
\ref{fig:var_cap_expon_channel} as a function of the delay.

\subsection{Interdependence of spatial and temporal correlations}
\label{sec:spatiotemporal interdependence}

In the previous section, we analyzed the situation where all paths
had the same transmission correlation matrices $\vec T_l=\vec
I_\nt$ resulting to significant simplifications. This situation is
not necessarily realistic. Typically, each path has an angle
spread smaller than the composite angle-spread and with a
different mean angle of departure from the transmitter for each
path.\cite{SCM_3GPP_TR25_996} Thus, even if the composite
narrowband correlations at the transmitter are assumed to be low,
the associated correlations per path may be substantial. It is
therefore interesting to compare the mutual information
distribution of the following two situations: In the first, all
paths have a correlation matrix identical to the narrowband
composite correlation matrix. In the second, each path has
different correlation matrices, subject to giving the same
narrowband correlation matrix as in the first case. For simplicity
we will take the narrowband composite correlation matrix to be
unity, with the following correlations between transmitting
antennas:
\begin{equation}
\label{eq:spatialcorrelations}
    T_{ab} = \int_{-180}^{180} \frac{d\phi}{\sqrt{2 \pi \delta^2}} e^{2
    \pi i (a - b) d_\lambda \sin((\phi+\phi_0)\pi/180) - \phi^2/(2 \delta^2)}
\end{equation}
with $a,b= 1 \ldots \nt$ being the index of transmitting antennas.
This is a simple model for the antenna correlations of a uniform
linear ideal antenna array with $ d_\lambda = d_{min}/\lambda$ the
nearest neighbor antenna spacing in wavelengths, a Gaussian power
azimuth spectrum with angle-spread $\delta$ degrees and $\phi_0$
degrees mean angle of departure.
\cite{Chizhik2000_CorrelatedMIMO1, Buehrer2000_CorrelationModel1}

\begin{figure}[htb]
\centerline{\epsfxsize=1.0\columnwidth\epsffile{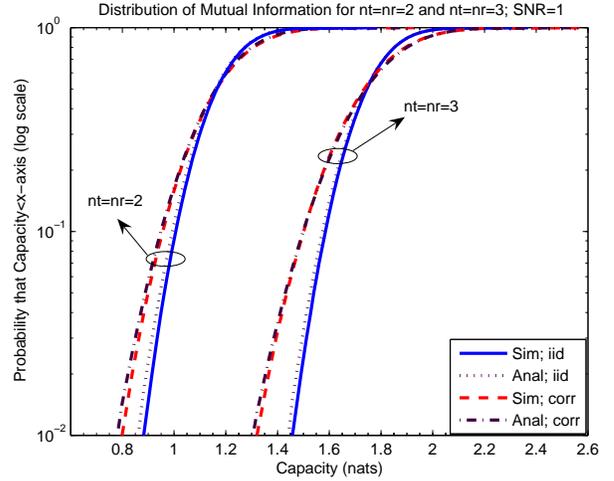}}
\caption{Cumulative distributions (CDF) of the mutual information
with two and three antenna arrays for signal-to-noise ratio (SNR)
$\rho=1$. 10 paths were used, each with an angle-spread of 18
degrees, with the mean angle of arrival of the $l$-th path
pointing at $18(l+1/2)$ degrees. While the mean mutual information
is nearly the same for both correlated and iid cases ($1.74$ nats
for $\nt=3$ and $1.16$ nats for $\nt=2$), the variance of the
correlated systems is nearly double the variance of the
corresponding iid case ($0.357$ vs. $0.0171$ for $\nt=3$ and
$0.0274$ vs. $0.0171$ for the $\nt=2$ case). The agreement between
the analytic large $N$ expression and the simulation is very good
down to $1\%$ outage.} \label{fig:cap_cdf_SNR1}
\end{figure}

In Fig. \ref{fig:cap_cdf_SNR1} we see that, although the mean
mutual information is identical in the cases, the variance of the
mutual information of the second case is roughly double to that of
the first case. We thus see that the correlation structure of the
underlying paths have a significant effect on the mutual
information distribution.

\subsection{Example: L distinct fully correlated paths}
\label{sec:distinct_fully correlated_paths}

As a final example, we describe a simple version of the general
non-Kronecker channel case given by (\ref{eq:G_lG_l*_gen_corrs}).
In particular, we assume that $\nt=\nr=n$ and that the correlation
matrices $\vec T_l$, $\vec R_l$ are mutually orthogonal, rank-one
matrices, e.g., for the transmitter we have $\vec T_l = n \vec a_l
\vec a_l^\dagger$, with $\vec a_l^\dagger \vec a_{l'} =
\delta_{ll'}$. This corresponds to
 a set of $L\leq n$
orthogonal plane-waves at the transmitter, each of which are
connected with a plane-wave arriving at the receiver in orthogonal
directions. In this case, (\ref{eq:r_gen_case}) and
(\ref{eq:t_gen_case}) simplify to
\begin{eqnarray}
\label{eq:r_orthogonal_plane_waves} %
 r_l &=& \frac{\rho_l q_l}{1+nq_l\rho_lt_l} \\
 \label{eq:t_orthogonal_plane_waves} %
 t_l &=& \frac{1}{1+nr_l}
\end{eqnarray}
where $q_l$ are the $L$ eigenvalues of $\vec Q$, given by
(\ref{eq:qeigs}). Assuming for simplicity that the $\rho_l$ are
ordered, i.e. $\rho_1\geq\rho_2\geq\ldots\rho_L$, the final
solution for the capacity-achieving input distribution covariance
matrix is
\begin{eqnarray}
\label{eq:Q_cap_achieving_orthogonal_plane_waves} %
\vec Q &=& \frac{1}{n} \sum_{l=1}^{L_{eff}} q_l \vec T_l \\
\label{eq:q_l_orthogonal_plane_waves} %
q_l &=& \frac{1}{\Lambda_{L_{eff}}} -
\frac{1}{\sqrt{\Lambda_{L_{eff}}}n\rho_l}
\end{eqnarray}
where
\begin{eqnarray}
\label{eq:Lambda_L_eff_orthogonal_plane_waves} %
\Lambda_{m} &=&
\frac{\left(\sqrt{\alpha_m^2+4m}-\alpha_m\right)^2}{4n} \\
\nonumber %
 \alpha_m &=&
\frac{1}{n}\sum_{l=1}^m\frac{1}{\sqrt{\rho_l}}
\end{eqnarray}
Here, $L_{eff}$ is the number of non-zero $\vec Q$ eigenvalues,
chosen with the condition
\begin{eqnarray}
\label{eq:L_eff_condition_orthogonal_plane_waves} %
\Lambda_{L_{eff}}< n\rho_l
\end{eqnarray}
for all $l\leq L_{eff}$, which comes from the requirement $r_l\geq
0$. The resulting ergodic capacity is
\begin{eqnarray}
\label{eq:ergod_capacity_orthogonal_plane_waves} %
I = \sum_{l=1}^{L_{eff}}
\left[\log\frac{n\rho_l}{\Lambda_{L_{eff}}} -
n\left(1-\sqrt{\frac{\Lambda_{L_{eff}}}{n\rho_l}}\right)\right]
\end{eqnarray}
From (\ref{eq:T_tilde}) and
(\ref{eq:Q_cap_achieving_orthogonal_plane_waves}), we see that the
capacity-achieving covariance matrix is a non-trivial linear
combination of $\vec T_l$,  each with coefficient $t_l\rho_l$,
which is obtained by solving (\ref{eq:r_orthogonal_plane_waves}),
(\ref{eq:t_orthogonal_plane_waves}) and
(\ref{eq:q_l_orthogonal_plane_waves}), which depends on the
properties of all paths.

\section{Conclusion}
\label{sec:conclusion}

In conclusion, we have presented an analytic approach to calculate
the statistics of the mutual information of MIMO systems for the
most general zero-mean Gaussian wideband channels. We have also
shown how the ergodic capacity can be calculated by optimizing
over the Gaussian input signal distribution. The analytic approach
is in principle valid for large antenna numbers, in which limit
the mutual information distribution approaches a Gaussian,
irrespective of the wideband richness of the channel. Thus the
outage capacity can be explicitly calculated. Nevertheless all
results have been found numerically to be valid with high accuracy
to arrays with few antennas. Thus our results are applicable to a
wide range of multipath problems, including, but not limited to,
multipath channels with a few delay taps or to an arbitrary
continuous power-delay profile and dual-polarized antennas with
arbitrary correlations. It should also be noted that this method
generalizes the so-called IND separable channels analyzed in
\cite{Tulino2004_INDLargeNCapacity} to general non-separable IND
 channels with arbitrary non-Kronecker product correlations.

This analytic approach provides the framework and a simple tool to
accurately analyze the statistics of throughput of even small
arrays in the presence of arbitrary channel correlations.

\appendices
\section{Complex Integrals}
\label{app:complex_integrals}

\begin{identity}
\label{id:completing_sq1}%
Let $\vec X, \, \vec A, \, \vec B$ be respectively $m\times n$
complex matrices and $\vec N, \, \vec M$ positive-definite
hermitian $n \times n$ and $m\times m$. Then, the following
equality holds
\begin{eqnarray}
\label{eq:complex_complete_square} %
&&\left(\det{[\vec N \otimes \vec M]}\right)^{-1} e^{-\frac{1}{2}
\Tr{\vec N^{-1} \vec A^\dagger \vec M^{-1} \vec B}}
\\
& & ~~~~~~ = \int D\vec X \, \, e^{-\frac{1}{2} \Tr{\vec N \vec
X^\dagger \vec M \vec X + \vec A^\dagger \vec X -\vec X^\dagger
\vec B }} \nonumber
\end{eqnarray}
where the integration measure $D\vec X$ is given by
(\ref{eq:DcX}).

\begin{proof}
See Appendix I in \cite{Moustakas2003_MIMO1}. Note that this
formula was printed incorrectly in that reference. Here we state
the relevant identity.

\end{proof}
There are several useful special cases of this identity.  Setting,
$\vec A = \vec 0$ and $\vec B = \vec 0$, we obtain
\begin{eqnarray}
\label{eq:complex_det_nu0} %
\left(\det{[\vec N \otimes \vec M]}\right)^{-1}  &=& (\det \vec
N)^{-m} (\det \vec M)^{-n}
\\ \nonumber &=& \int D\vec X \,\,
e^{-\frac{1}{2} \Tr{\vec N \vec X^\dagger \vec M \vec X}}
\end{eqnarray}
Further setting $\vec N=\vec I_{n}$ yields
\begin{equation}
\label{eq:complex_det_nu} %
 (\det \vec M)^{-n} = \int D\vec X \,\, e^{-\frac{1}{2}
\Tr{\vec X^\dagger \vec M \vec X}}
\end{equation}
\end{identity}

\begin{identity}[Hubbard-Stratonovich Transformation]
\label{id:hub_strat}%
Let $\vec U$, $\vec V$ be arbitrary complex $M \nu\times M \nu$
matrices, where $\nu$ is assumed to be an arbitrary positive
integer. Then the following identity holds
\begin{equation}
\label{eq:hub_strat_identity}%
e^{-\mbox{\small Tr}[\vec U\vec V]} = \int d\mu({\cal  T}, {\cal
R}) e^{\mbox{\small Tr}[{\cal R  T}-  \vec U {\cal  T} - {\cal R}
\vec V ]}
\end{equation}
In the above equation, the auxiliary matrices $\cal T$ and $\cal
R$ are general complex matrices $M \nu \times M \nu$ and their
integration measure is given by (\ref{eq:dmuTR1}). The integration
of the elements of $\cal R$ and $\cal T$ is along contours in
complex space parallel to the real and imaginary axis respectively
as discussed in \cite{Moustakas2003_MIMO1}.
\end{identity}
\begin{proof}
See Appendix I in \cite{Moustakas2003_MIMO1}.
\end{proof}

\section{Derivation of (\ref{eq:gnu_S}), (\ref{eq:S_full})}
\label{app:Derivation of gnu_S}
 In this Appendix we will express
$g(\nu)$ as in (\ref{eq:gnu_S}), (\ref{eq:S_full}). We start with
(\ref{eq:g_nu_def}) assuming that $\nu$ is an arbitrary positive
integer.  Using (\ref{eq:complex_det_nu}) we can write
\begin{eqnarray}
\label{eq:Xint}
  \det {\mbox{\Large$($}} \vec I_\nr \!\!\!\!\! &+& \!\!\!\!\! \bGhat_k \vec Q
 \bGhat_k{}^\dagger {\mbox{\Large$)$}}^{-\nu}     \\
 &=& \int D\vec X_k   e^{-\frac{1}{2}\Tr{\vec X_k^\dagger \vec
 X_k^{\phantom{\dagger}} + \vec X_k^\dagger \bGhat_k^{\phantom{\dagger}} \vec Q
 \bGhat_k{}^\dagger  \vec
 X_k^{\phantom{\dagger}}}}\nonumber
\end{eqnarray}
where $\vec X_k$ is an $\nt\times\nu$-dimensional complex matrix.
We then further use (\ref{eq:complex_complete_square}) to write
\begin{eqnarray} \label{eq:Yint}
   & &  e^{-\frac{1}{2}\Tr{\vec X_k^\dagger \bGhat_k^{\phantom{\dagger}} \vec Q
 \bGhat_k{}^\dagger  \vec
 X_k^{\phantom{\dagger}}}} \\  &=& \int D\vec Y_k  e^{-\frac{1}{2} \Tr{\vec Y_k^\dagger \vec Y_k^{\phantom{\dagger}}
 +  \vec X_k^\dagger \bGhat_k^{\phantom{\dagger}} \vec Q^{1/2} \vec Y_k^{\phantom{\dagger}}
 - \vec Y_k^\dagger \vec Q^{1/2} \bGhat_k{}^\dagger  \vec
 X_k^{\phantom{\dagger}}}} \nonumber
\end{eqnarray}
where $\vec Y_k$ is also an $\nt\times\nu$-dimensional complex
matrix. Thus, using (\ref{eq:Xint}) and (\ref{eq:Yint}) and the
definition (\ref{eq:g_nu_def}) of $g(\nu)$ we can write
\begin{eqnarray}
\label{eq:g_nu_XYG} \nonumber g(\nu)&=&   \left<  \prod_k\int
D\vec X_k\, D\vec Y_k\, e^{-\frac{1}{2}
\sum_k\Tr{\vec X_k^\dagger \vec X_k^{\phantom{\dagger}} + \vec Y_k^\dagger \vec Y_k^{\phantom{\dagger}}}} \right. \\
 && \left.  e^{-\frac{1}{2} \sum_k\Tr{\vec X_k^\dagger \bGhat_k^{\phantom{\dagger}}
\vec Q^{1/2}\vec Y_k^{\phantom{\dagger}} - \vec Y_k^\dagger \vec
Q^{1/2} \bGhat_k{}^\dagger \vec X_k^{\phantom{\dagger}}} }
\rule[15pt]{0pt}{0pt} \right>
\end{eqnarray}
where $k$ ranges from $0$ to $M-1$.  Note that, as discussed
above, we have been able to set all $\vec Q_k$ equal to a single
$\vec Q$.

To average the bracketed term over channel realizations we use
(\ref{eq:G_k_ft_def}) to express $\bGhat$ in terms of $\bG_l$. The
probability density of $\bG_l$ is defined by
(\ref{eq:G_lG_l*_gen_corrs}) and can be rewritten explicitly
\begin{equation}\label{eq:pdf_G_l}
  p(\bG_l)  = \det\left[\frac{\rho_l}{\nt}\vec T_l \otimes \vec R_l\right]^{-1}
   \!\!\! e^{-\frac{\nt}{2 \rho_l} \Tr{\vec T^{-1}_l
  \bG_l^\dagger \vec R^{-1}_l \bG_l^{\phantom{\dagger}}}}
\end{equation}
The expectation bracket of any operator $F(\{\bG_l\})$ which is a
function of the $\bG_l$'s can then be written as
\begin{equation}
\label{eq:pdfint}
    \left\langle  \rule[11pt]{0pt}{0pt} F(\{\bG_l\}) \right\rangle = \prod_{l=0}^{L-1} \int D\bG_l \, p(\bG_l)  \,\,\,\,
    F(\{\bG_l\})
\end{equation}
Note that using (\ref{eq:complex_det_nu0}) it is easy to see that
this probability distribution is properly normalized (i.e.,
$\langle 1 \rangle = 1$).

We now evaluate the expectation bracket in (\ref{eq:g_nu_XYG}) by
rewriting $\bGhat$ in terms of $\bG_l$ and integrating over the
channel realizations (using (\ref{eq:pdf_G_l}) and
(\ref{eq:pdfint} and applying (\ref{eq:complex_complete_square})
to perform the integral). As a result we obtain
\begin{eqnarray}
\label{eq:brack_quant_g_nu} %
& &  g(\nu)=    \prod_k\int D\vec X_k\, D\vec Y_k\,
e^{-\frac{1}{2}
\sum_k\Tr{\vec X_k^\dagger \vec X_k^{\phantom{\dagger}} + \vec Y_k^\dagger \vec Y_k^{\phantom{\dagger}}}} \\
 & & \cdot
\prod_{l} e^{-\left[\frac{\rho_l}{2\nt} \sum_{kk'}e^{\frac{2\pi i
(k-k')m_l}{M}} \Tr{  \vec X_{k'}^\dagger \vec R_l \vec
X_k^{\phantom{\dagger}} \vec Y_k^\dagger \vec Q^{1/2} \vec T_l
\vec Q^{1/2} \vec Y_{k'}^{\phantom{\dagger}}}\right]} \nonumber
\end{eqnarray}
Following \cite{Moustakas2003_MIMO1} we use Identity
\ref{id:hub_strat} in Appendix \ref{app:complex_integrals} to
express the above in a quadratic form in terms of $\vec X_k$,
$\vec Y_k$ by introducing $2L$ $M\nu\times M\nu$ matrices
$\bRcal^l$, $\bTcal^l$. These matrices, whenever convenient, will
be represented $\bRcal^l_{kk'}$, $\bTcal^l_{kk'}$, as a set of $L
M^2$ matrices of dimension $\nu\times\nu$ each. Thus the second
line of (\ref{eq:brack_quant_g_nu}) becomes
\begin{eqnarray}
\label{eq:g_nu_TR1} %
&&\int d\mu(\{\bTcal^l, \bRcal^l\}) \prod_l\prod_{kk'}
\left(\exp\left[\Tr{\bTcal_{kk'}^{l} \bRcal_{kk'}^{l}}\right]
\right. \\ \nonumber %
&&\cdot \left.\exp\left[-\frac{\rho_l}{2\sqrt{\nt}}e^{\frac{2\pi
i(k-k')m_l}{M}} \Tr{\bTcal^l_{k'k} \vec Y_k^\dagger \vec
Q^{1/2}\vec T_l \vec Q^{1/2}\vec
Y_{k'}}  \right] \right.\\ \nonumber %
&&\cdot \left. \exp\left[-\frac{1}{2\sqrt{\nt}}\Tr{\vec
X_{k'}^\dagger \vec R_l \vec X_k \bRcal^l_{kk'}}\right] \right)
\end{eqnarray}
Combining  (\ref{eq:brack_quant_g_nu}) and (\ref{eq:g_nu_TR1}) and
using  (\ref{eq:complex_det_nu0}), we can now integrate over $\vec
X_k$, $\vec Y_k$, resulting in
\begin{equation} \label{eq:gnu_S_app}
   g(\nu)  =
  \int d\mu(\{\bTcal^l, \bRcal^l\})  e^{-{\cal S }}
\end{equation}
with ${\cal S}$ given in (\ref{eq:S_full}).

\section{Details for Saddle Point Analysis of (\ref{eq:gnu_S}), (\ref{eq:S_full})}
\label{app:Details for Saddle Point Analysis of gnu_S}

Using the change of variables $\bTcal^l \rightarrow \dTcal^l$,
$\bRcal^l \rightarrow \dRcal^l$ defined in
(\ref{eq:TR_saddle_point_vicinity}) we expand ${\cal S}$ in
(\ref{eq:S_full}) in powers of $\dTcal^l$, $\dRcal^l$, resulting
in
\begin{eqnarray}
\label{eq:expanded_S} %
\label{eq:saddle0}
    {\cal S}_0 &=&  \nu \left[
     M \log \det \left( \vec I_\nt + \,
      \sum_l t_l \rho_l \vec Q \vec T_l   \right) \right.
        \\ \nonumber
       &+& \left. M\log \det \left(\vec I_\nr + \sum_l r_l \vec R_l   \right)
       - \nt M \sum_l r_l t_l   \right] \\ \nonumber
    \\ \label{eq:S1_expanded}%
    {\cal S}_1 &=& \sum_{k,l}\left[ \left(M^l_{r,1} - t_l\sqrt{\nt}\right)
    \Tr{\dRcal^l_{kk}} \right.\\ \nonumber
    &&+ \left.\left(M^l_{t,1} - r_l\sqrt{\nt} \right)
    \Tr{\dTcal^l_{kk}}\right]
    \\ \nonumber
    {\cal S}_2 &=& -\frac{1}{2} \sum_{kk'}\sum_{ll'}
\mbox{Tr}\left\{M^{ll'}_{r,2} \,\dRcal^l_{kk'}\dRcal^{l'}_{k'k}
\right.
\\ \nonumber
         &&\left. +  M^{ll'}_{t,2} \,   \dTcal^l_{kk'}\dTcal^{l'}_{k'k} +  2 \dTcal^l_{kk'}
         \dRcal^{l'}_{k'k} \right\}
    \\ \label{eq:S2_expanded}%
        &=& \frac{1}{2}  \sum_{kk'}\Tr{\vec x_{kk'} {\bf V}^{kk'}
        \vec x_{k'k}^T}
\end{eqnarray}
where the $2L$-dimensional vector $\vec x_{kk'}$ of $\nu\times\nu$
matrices is defined as
\begin{equation}
\label{eq:x_qq'_vec_def}%
 \vec x_{kk'} =  \left[ \dRcal^0_{kk'} \ldots \dRcal^L_{kk'} \dTcal^0_{kk'} \ldots \dTcal^L_{kk'} \right]
\end{equation}
and the corresponding $2L$-dimensional Hessian $\vec V^{kk'}$ is
expressed in block-diagonal form as
\begin{equation}
\label{eq:hessian}%
 {\bf V}^{kk'} =  \left[ \begin{array}{cc}
-\vec M_{r,2} & -\vec I_L
        \\ -\vec I_L &  -\vec M_{t,2} \end{array} \right]
\end{equation}
where the matrices $\vec M_{r,2}$, $\vec M_{t,2}$ in the diagonals
have elements $M^{ll'}_{r,2}$ and  $M^{ll'}_{t,2}$, respectively,
with $l=0,\ldots,L-1$. For $p
> 2$ the expanded terms take the form
\begin{eqnarray}
    \label{eq:S_p_greater2}
    {\cal S}_p &=& \frac{(-1)^p}{p} \sum_{\vec k_p,\vec l_p}\left[  M^{\vec l_p}_{r,p}\,
    \mbox{Tr} \left\{
    \dRcal^{l_1}_{k_1k_2}\cdots\dRcal^{l_p}_{k_pk_1}\right\} \right.\\ \nonumber
    &&\left. + M^{\vec l_p}_{t,p} \, \mbox{Tr} \left\{ \dTcal^{l_1}_{k_1k_2}\cdots\dTcal^{l_p}_{k_pk_1}
    \right\}\right]
\end{eqnarray}
where the $p$-dimensional integer valued vectors $\vec
l=[l_1\ldots l_p]$, $\vec k_p=[k_1\ldots k_p]$ are being summed
over. The coefficients in this Taylor expansion have the form
\begin{eqnarray} \label{eq:Mdef}
 M^{\vec l_p}_{t,p} &=&
  \mbox{Tr}\left\{\prod_{i=1}^p\left[
    \left(\vec I_\nt + \vec Q\sum_l t_l \rho_l  \vec T_l \right)^{-1}
    \right.\right.
    \\ \nonumber %
    &&\left.\left. \frac{\rho_{l_i}\vec Q \vec
    T_{l_i} e^{\frac{2\pi i(k_i-k_{i+1})m_{l_i}}{M}}}{\sqrt{\nt}}
    \right] \right\}\nonumber \\ \label{eq:Mdef2}
    M^{\vec l_p}_{r,p} &=& \Tr{\prod_{i=1}^p\left[
    \left(\vec I_\nr + \sum_l r_l \vec R_l \right)^{-1} \frac{\vec R_{l_i}}{\sqrt{\nt}}
    \right]}
\end{eqnarray}
Note that while in (\ref{eq:g_nu_TR1}) $\vec Q$ appears in the
form $\vec Q^{1/2}\vec T_l \vec Q^{1/2}$, in (\ref{eq:saddle0}),
(\ref{eq:Mdef}) it is possible to combine the two $\vec Q^{1/2}$
into a single $\vec Q$.

\section{Higher Order Terms}
\label{app:higher_order_terms}

In this section we will follow the formulation of
\cite{Moustakas2003_MIMO1} to calculate the leading $1/n$
correction to $g(\nu)$ which will contribute as a leading term to
the skewness ${\cal C}_3$ and as a correction to the average
mutual information ${\cal C}_1$.

We define an expectation bracket of $F(\dTcal,\dRcal)$, an
arbitrary function of $\dTcal$, $\dRcal$, as
\begin{equation}
\label{eq:double_bracket_def}
 \langle\langle F \rangle \rangle = \prod_{kk'}\left|\det\vec V^{kk'}\right|^{\nu^2/2} \int d \mu\left(\dTcal, \, \dRcal
 \right) e^{-{\cal S}_2} F(\dTcal, \, \dRcal)
\end{equation}
To calculate such expectations, we will expand the function $F$ in
its arguments and will then integrate over the Gaussian integral.
Thus only integrals over even powers of $\dTcal$, $\dRcal$ will
survive. To evaluate the expanded terms we need the following
second order moments (see below)
\begin{eqnarray}\label{eq:quadratic_corrs}%
    \langle \langle \delta R^{p}_{k_1k_2,ab} \delta R^{q}_{k_3k_4,cd} \rangle
    \rangle &=& - \delta_{k_1k_4} \delta_{k_2k_3} \delta_{ad} \delta_{bc} W_{1,pq}^{k_1k_2}  \nonumber \\
    \langle \langle \delta T^{p}_{k_1k_2,ab} \delta T^{q}_{k_3k_4,cd} \rangle
    \rangle &=& - \delta_{k_1k_4} \delta_{k_2k_3} \delta_{ad} \delta_{bc} W_{2,pq}^{k_1k_2} \nonumber \\
    \langle \langle \delta T^{p}_{k_1k_2,ab} \delta R^{q}_{k_3k_4,cd} \rangle
    \rangle &=& - \delta_{k_1k_4} \delta_{k_2k_3} \delta_{ad} \delta_{bc} W_{3,pq}^{k_1k_2} \nonumber \\
\end{eqnarray}
where for each $k_1, k_2 = 1, \ldots, \nu$, the $L\times L$
matrices $\vec W_{i}^{k_1k_2}$ for $i=1,\ldots,3$ are given in
terms of the $L\times L$ matrices $\vec M_{r,2}$, $\vec M_{t,2}$
(see (\ref{eq:Mdef}), (\ref{eq:Mdef2})) by the following
expressions
\begin{eqnarray}\label{eq:quadratic_corrs_W_matrices}%
    \vec W_{1}^{k_1k_2} &=& - \vec M_{t,2} \left[ \vec M_{r,2}\vec M_{t,2}- \vec I_L\right]^{-1} \\ \nonumber
    \vec W_{2}^{k_1k_2} &=& - \vec M_{r,2} \left[ \vec M_{t,2}\vec M_{r,2}- \vec I_L\right]^{-1} \\ \nonumber
    \vec W_{3}^{k_1k_2} &=& \left[ \vec M_{r,2}\vec M_{t,2}- \vec I_L\right]^{-1}
\end{eqnarray}
independent of $k_1, k_2$. In our particular case, the function
$F(\dTcal,\dRcal)$ is $\exp[-\sum_{p>2} {\cal S}_p]$, with ${\cal
S}_p$ expressed in terms of $\dTcal$, $\dRcal$, as in
(\ref{eq:S_p_greater2}). We now expand the exponential by
combining terms with equal powers of $n$. To do this, we note in
(\ref{eq:S_p_greater2}) that $\langle \langle {\cal S}_p \rangle
\rangle$ is of order $n^{-p/2 +1}$ for $p$ even, while it is zero
for $p$ odd. Keeping only the ${\cal O}(n^{-1})$ terms, $g(\nu)$
takes the form
\begin{equation}
  g(\nu) =  e^{-\nu M \Gamma} \prod_{pq}\left| \det {\bf
    V}^{pq} \right|^{-\nu^2/2} \left[ 1 + D_1 + {\cal
    O}(n^{-2})\right] \label{eq:Dexpansion_app}
\end{equation}
where
\begin{eqnarray}
 D_1 = \langle \langle {\cal S}_4  + \frac{1}{2} {\cal S}_3^2 \rangle \rangle
\end{eqnarray}
which is of order $1/n$.

To evaluate $D_1$ we need to calculate $\langle\langle {\cal
S}_4\rangle\rangle $ and $\langle\langle {\cal
S}_3^2\rangle\rangle$, which, as seen in (\ref{eq:S_p_greater2}),
include fourth order and sixth order products in $\dTcal$,
$\dRcal$, respectively. These can be calculated by applying Wick's
theorem (see \cite{Moustakas2003_MIMO1} or
\cite{Reed1962_WickLikeTheorem}), i.e. by ``pairing'' all
$\dTcal$'s and $\dRcal$'s with each other and using
(\ref{eq:quadratic_corrs}) to calculate the corresponding
quadratic moments. As an example, we evaluate below the term in
${\cal S}_4$ which is proportional to $M^{p_1p_2p_3p_4}_{r,4}$ in
(\ref{eq:S_p_greater2}).
\begin{eqnarray}\label{eq:<<R4>>_line1}
&&  \sum_{k_1,\ldots,k_4=1}^M\sum_{a,b,c,d=1}^\nu  \langle \langle
\delta R^{p_1}_{k_1k_2,ab} \ldots \delta R^{p_4}_{k_4k_1,da}
\rangle \rangle
    \\ \nonumber%
    &=& \sum_{k_1,\ldots,k_4=1}^M\sum_{a,b,c,d=1}^\nu  \left[ \right. \\  \nonumber %
    && \left.\langle \langle \delta R^{p_1}_{k_1k_2,ab}
    \delta R^{p_2}_{k_2k_3,bc}
    \rangle \rangle %
    \langle \langle \delta R^{p_3}_{k_3k_4,cd}\delta  R^{p_4}_{k_4k_1,da} %
    \rangle \rangle \right. \\ \nonumber
    &&+ \left. %
    \langle \langle \delta R^{p_1}_{k_1k_2,ab}
    \delta R^{p_3}_{k_3k_4,cd}
    \rangle \rangle %
    \langle \langle \delta R^{p_2}_{k_2k_3,bc}\delta  R^{p_4}_{k_4k_1,da} %
    \rangle \rangle \right. \\ \nonumber
    &&+ \left. %
    \langle \langle \delta R^{p_1}_{k_1k_2,ab}
    \delta R^{p_4}_{k_4k_1,da}
    \rangle \rangle %
    \langle \langle \delta R^{p_2}_{k_2k_3,bc}\delta  R^{p_3}_{k_3k_4,cd} %
    \rangle \rangle \right]
    \\ \nonumber
    &=&  \nu^3 \left(W_{1,p_1p_3}^{k_1k_2} W_{1,p_3p_4}^{k_1k_3} + W_{1,p_1p_4}^{k_1k_2} W_{1,p_2p_3}^{k_2k_3} \right)%
    \\ \label{eq:<<R4>>_line3} %
    &&+ \nu M W_{1,p_1p_3}^{k_1k_1} W_{1,p_2p_4}^{k_1k_1}
\end{eqnarray}
We can similarly evaluate the second term in ${\cal S}_4$ as well
as ${\cal S}_3^2$ to get
\begin{equation}
    \label{eq:D_1_app}
    D_1 = a_1 \nu + a_3 \nu^3
\end{equation}
where
\begin{eqnarray}
\label{eq:a1def}
    a_1 &=& \sum_{p_1\ldots p_4}\left\{\frac{M}{4} \, \left[ M_{r,4}^{p_1p_2p_3p_4}W_{1,p_1p_3}^{kk} W_{1,p_2p_4}^{kk} \right.
    \right.\\ \nonumber %
    &&+ \left. M_{t,4}^{p_1p_2p_3p_4}W_{2,p_1p_3}^{kk} W_{2,p_2p_4}^{kk}\right]
    \\ \nonumber %
    &+& \sum_{p_1\ldots p_6}\left\{\frac{M}{6} \, \left[ M_{r,3}^{p_1p_2p_3} M_{r,3}^{p_4p_5p_6} W_{1,p_1p_2}^{kk} W_{1,p_3p_4}^{kk}
    W_{1,p_5p_6}^{kk} \right. \right.\\ \nonumber %
    &&+ \left. M_{t,3}^{p_1p_2p_3} M_{t,3}^{p_4p_5p_6} W_{2,p_1p_2}^{kk} W_{2,p_3p_4}^{kk}
    W_{2,p_5p_6}^{kk} \right. \\ \nonumber %
    &&+ \left.\left. 2M_{r,3}^{p_1p_2p_3} M_{t,3}^{p_4p_5p_6} W_{3,p_1p_4}^{kk} W_{3,p_2p_5}^{kk}
    W_{3,p_3p_6}^{kk} \right]\right\}
\end{eqnarray}
and
\begin{eqnarray}
\label{eq:a3def}
    a_3 &=&  \frac{1}{4}\sum_{p_1\ldots p_4}\sum_{k_1k_2k_3} \left\{ \right.\\ \nonumber%
&&\left.M^{p_1p_2p_3p_4}_{r,4}\left[W_{1,p_1p_2}^{k_1k_2}
W_{1,p_3p_4}^{k_1k_3} + W_{1,p_1p_4}^{k_1k_2}
W_{1,p_2p_3}^{k_2k_3}\right] \right.\\ \nonumber %
&&\left.+M^{p_1p_2p_3p_4}_{t,4}\left[W_{2,p_1p_2}^{k_1k_2}
W_{2,p_3p_4}^{k_1k_3} + W_{2,p_1p_4}^{k_1k_2}
W_{2,p_2p_3}^{k_2k_3}\right]\right\} \\ \nonumber %
&+&  \frac{1}{6}\sum_{p_1\ldots p_6}\sum_{k_1k_2k_3} \left\{ %
M_{r,3}^{p_1p_2p_3} M_{r,3}^{p_4p_5p_6}  \right.\\
\nonumber%
&&\left.\left[3W_{1,p_1p_2}^{k_1k_2}
W_{1,p_3p_4}^{k_1k_1}W_{1,p_5p_6}^{k_1k_3}
+W_{1,p_1p_4}^{k_1k_2} W_{1,p_2p_6}^{k_2k_3}W_{1,p_3p_5}^{k_3k_1}\right] \right.\\
\nonumber%
&&+\left.%
M_{t,3}^{p_1p_2p_3} M_{t,3}^{p_4p_5p_6} \right. \\
\nonumber%
&&\left.\left[ 3W_{2,p_1p_2}^{k_1k_2}
W_{2,p_3p_4}^{k_1k_1}W_{2,p_5p_6}^{k_1k_3}
+W_{2,p_1p_4}^{k_1k_2} W_{2,p_2p_6}^{k_2k_3}W_{2,p_3p_5}^{k_3k_1}\right]\right.\\
\nonumber%
&&+\left.%
2M_{r,3}^{p_1p_2p_3} M_{t,3}^{p_4p_5p_6} \right. \\
\nonumber%
&&\left.\left[ 3W_{1,p_1p_2}^{k_1k_2}
W_{2,p_5p_6}^{k_1k_3}W_{3,p_3p_4}^{k_1k_1}
+W_{3,p_1p_4}^{k_1k_2} W_{3,p_2p_6}^{k_2k_3}W_{3,p_3p_2}^{k_3k_1}\right]\right\}\\
\nonumber%
\end{eqnarray}

\section{Capacity-Achieving Input Signal Covariance $\vec Q$}
\label{app:optimalQ}

In this Appendix we will show that the capacity-achieving input
distribution $\vec Q$ is diagonal in the basis of $\tilde {\bf T}$
defined in (\ref{eq:T_tilde}). To start the proof we point out
that the mutual information (to order ${\cal O}(1/n)$) for a given
$\vec Q$ is the extremum of
(\ref{eq:ergodic_mutual_information_1storder}) given also below
\begin{eqnarray}
 \Gamma(\vec Q, \{ t_l\}, \{ r_l\}) &=&
     M \log \det \left( \vec I_\nt + \,
      \vec Q {\tilde {\bf T}}   \right) \nonumber
        \\ \nonumber
       &+& M\log \det \left(\vec I_\nr + {\tilde {\bf R}}  \right)  - \nt M \sum_l r_l t_l
\end{eqnarray}
with respect to $t_l$, $r_l$ for $l=0,\ldots,L-1$. The
saddle-point equations are given by (\ref{eq:r_gen_case}),
(\ref{eq:t_gen_case}) also seen below:
\begin{eqnarray}
 r_l &=& \frac{\rho_l}{\nt} \Tr{ \vec Q \vec T_l \left[\vec I_{\nt} +
 \vec Q {\tilde {\bf T}} \right]^{-1}} \nonumber
  \\ \nonumber
  t_l &=& \frac{1}{\nt} \Tr{\vec R_l \left[\vec I_\nr+{\tilde {\bf R}} \right]^{ -1}  }
\end{eqnarray}
It should be noted that the mutual information is an extremum of
${\cal S}$ in a larger complex space of the elements of the
matrices $\{\bTcal^l, \bRcal^l\}$, but for simplicity we only
focus on the dependence of $\Gamma(\vec Q, \{ t_l\}, \{ r_l\})$ in
the $2L$-dimensional space of $\{ t_l,  r_l\}$. In this case one
can view $\Gamma(\vec Q, \{ t_l\}, \{ r_l\})$ as a function of
$2L+\nt^2+\nt-1$ variables, where the last $\nt^2+\nt-1$ are the
degrees of freedom of $\vec Q$, an $\nt$-dimensional hermitian
complex matrix with fixed trace. Extremizing the above function
over $\{ t_l, r_l\}$ we can eliminate all $\{ t_l, r_l\}$ using
the above equations. Thus for fixed $\vec Q$ the mutual
information can be written as $I(\vec Q) = \Gamma(\vec Q,
\{t_l(\vec Q)\}, \{r_l(\vec Q)\})$ with $t_l(\vec Q), r_l(\vec Q)$
functions of $\vec Q$. Suppose now that we maximize $I(\vec Q)$
with respect to $\vec Q$ with the constraint $Tr\vec Q=\nt$ and
that $\vec Q_0$ is the optimal matrix. As a result, $I(\vec
Q_0)=\Gamma(\vec Q_0, \{t_l(\vec Q_0)\}, \{r_l(\vec Q_0)\})$ is a
maximum over $\vec Q$ and an extremum over $t_l, r_l$. Thus if one
varies $\vec Q_0$ locally keeping its eigenvalues (and trace)
fixed, the variation of $I(\vec Q)$ will vanish to first order in
the variation. The most general such variation can be written as
\begin{eqnarray}
    \vec Q_\lambda &=& e^{i \lambda \vec H} {\bf Q}_0 e^{-i \lambda \vec H}
    \\
    &=& {\bf Q}_0 + i \lambda [\vec H, {\bf Q}_0
    ]  + \ldots
\end{eqnarray}
where $\vec H = \vec H^\dagger$ is an arbitrary traceless
Hermitian matrix, $\lambda$ is a small scalar, and the notation
$[a,b] = ab - ba$ is the commutator. Thus the first derivative of
$\Gamma$ with $\lambda$ has to vanish at $\lambda=0$. Therefore we
have
\begin{eqnarray}
   0 =  \left.\frac{d \Gamma}{d \lambda}\right|_{\lambda = 0} =
   \left.\left[\frac{\partial \Gamma}{\partial \lambda}
   + \sum_l \left\{ \frac{\partial \Gamma}{\partial t_l} \frac{d t_l}{d \lambda}
   + \frac{\partial \Gamma}{\partial t_l} \frac{d t_l}{d
   \lambda}\right\}\right]\right|_{\lambda =
   0}
\end{eqnarray}
Since $\Gamma$ is an extremum with respect to $\{r_l,t_l\}$ the
partial derivatives of $\Gamma$ with $\{r_l,t_l\}$ vanish. We are
left with the first term, $\partial \Gamma/\partial\lambda$, which
should also vanish if $\Gamma$ is a maximum over $\vec Q$,
resulting to
\begin{eqnarray}
   0 = \left.\frac{\partial \Gamma}{\partial \lambda}\right|_{\lambda=0} %
   &=& \mbox{Tr}\left[ [\vec H,{\bf Q}_0] \tilde {\bf T} \left(\vec I_{n_t} +
{\bf Q}_0 \tilde {\bf T}\right)^{-1}  \right]  \\
&=& \mbox{Tr} [\vec H \vec Z] \label{eq:HZ}
\end{eqnarray}
with
\begin{eqnarray}
\label{eq:Zdef} \vec Z &=&  \left(\vec I_{n_t} + {\bf Q}_0 \tilde
{\bf T}\right)^{-1} -\left(\vec I_{n_t} +  \tilde {\bf T} {\bf
Q}_0\right)^{-1}
\end{eqnarray}
Now, since $\vec H$ is an arbitrary traceless Hermitian matrix,
the condition (\ref{eq:HZ}) is equivalent to the statement that
$\vec Z$ is proportional to the identity matrix.  However, it is
easy to see from (\ref{eq:Zdef}) that $\vec Z$ must be traceless,
which implies that our extremization condition is equivalent to
$\vec Z = \vec 0$ or
\begin{equation}
\tilde {\bf T} {\bf Q}_0 = {\bf Q}_0  \tilde {\bf T}
\end{equation}
which requires that ${\bf Q}_0$ and $\tilde {\bf T}$ have the same
eigenvectors whenever ${\bf Q}_0$ is a maximum.


\end{document}